\documentclass[11pt,a4paper]{article}
\pdfoutput=1

\usepackage[]{Packages}

\RequirePackage{placeins}
\RequirePackage{ragged2e}
\usepackage{SimeonTeX}

\usepackage{Definitions}

\allowdisplaybreaks[1]

\preprint{IPMU11-0096}

\newcommand{\OfficialTitle}{String theory of the Omega deformation}

\title{\vspace{2cm}
  {\huge   \textbf{\OfficialTitle}}
}

\hypersetup{pdfauthor={Simeon Hellerman, Domenico Orlando and Susanne Reffert},pdftitle={\OfficialTitle}}

\author{
  \begin{minipage}{.8\linewidth}
    \vspace{1cm}
    \begin{center}
      {\small \textbf{Simeon Hellerman}, \textbf{Domenico Orlando} and
        \textbf{Susanne Reffert} }
    \end{center}
    \vspace{1cm}
    \begin{minipage}{\linewidth}\centering
      {\itshape \footnotesize 
        Institute for the Physics and Mathematics of
        the Universe, \\
        The University of Tokyo, Kashiwa-no-Ha 5-1-5, \\
        Kashiwa-shi, 277-8568 Chiba, Japan.\\
      }
    \end{minipage}
    \vspace{2cm}
    \begin{center}
      \itshape This paper is dedicated to the memory of the victims \\
      of the Tohoku earthquake and tsunami of March 2011. 
    \end{center}
  \end{minipage}
}

\date{}

\begin{document}

\setstretch{1.1}

\numberwithin{equation}{section}

\begin{titlepage}

  \maketitle

  \thispagestyle{empty}

  \vfill

  \abstract{\RaggedLeft In this article, we construct a supersymmetric real mass
    deformation for the adjoint chiral multiplets in the
    gauge theory describing the dynamics of a stack of D2--branes
    in type II string theory.  We do so by placing the D2--branes into
    the T--dual of a supersymmetric NS fluxbrane background. We
    furthermore note that this background is the string theoretic
    realization of an $\Omega$ deformation of flat space in the
    directions transverse to the branes where the deformation
    parameters satisfy $\varepsilon_1=-\varepsilon_2$. This $\Omega$
    deformation therefore serves to give supersymmetric real masses to the chiral
    multiplets of the 3D gauge theory.  To illustrate the physical effect of the real
    mass term, we derive \textsc{bps}-saturated classical 
    solutions for the branes rotating in this background.  Finally, we
    reproduce all the same structure in the presence of NS fivebranes
    and comment on the relationship to the gauge theory/spin-chain
    correspondence of Nekrasov and Shatashvili. }

\end{titlepage}

\clearpage
\section{Introduction}
\label{sec:introduction}

\FloatBarrier

In this article, we study the low energy effective gauge
theory describing the motions of a stack of D2--branes extended in the
$x^0,\, x^1,\, x^2$ directions. Our aim is to give SUSY-preserving real masses to the
fields describing the motions of the D2--branes in the directions $x^4,
\dots,x^7$. We will do so by placing the D2--branes into a closed
string background corresponding to the T--dual of a
supersymmetric NS fluxbrane~\cite{Melvin:1963qx,Tseytlin:1994ei,Tseytlin:1995zv,Russo:2001na,Gutperle:2001mb,Takayanagi:2001jj,Hellerman:2006tx}. We
will point out that the fluxbrane is the string theory realization of
an
$\Omega$--deformation~\cite{Moore:1997dj,Lossev:1997bz,Nekrasov:2002qd,Nekrasov:2003rj,Billo:2006jm,Nekrasov:2010ka}
of flat space in the directions $x^4, \dots,x^7$, where the
deformation parameters fulfill $\varepsilon_1=-\varepsilon_2$.

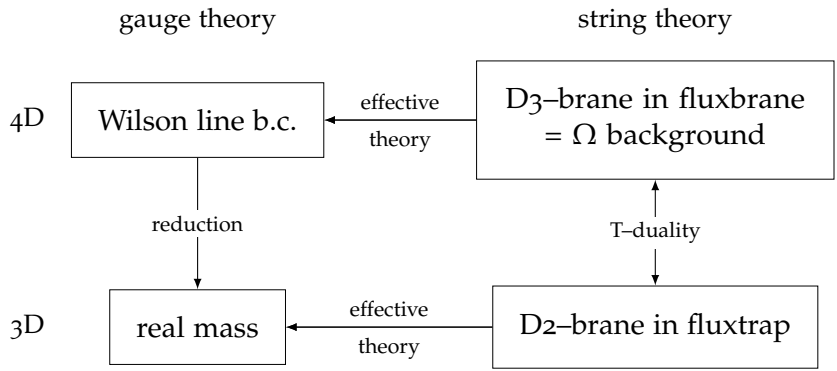
\begin{figure}
  \centering
  \begin{tikzpicture}[description/.style={fill=white,inner sep=2pt}]
    \matrix (m) [matrix of nodes, 
    nodes={
      draw,
      line width=.3pt,
      anchor=center, 
      text centered,
      inner sep=10pt,
    },
    txt/.style={text width=4cm,anchor=center},
    empty/.style={draw=none, minimum width=1cm,font=\small}]
    {  & |[empty]| gauge theory  &|[empty]|{} &|[empty]|{}& |[empty]| string theory  \\
      |[empty]| 4D & Wilson line b.c. &&& |[txt]| {D3--brane in fluxbrane \\= $\Omega$ background} \\
      |[empty]| {} \\ |[empty]| {} \\
      |[empty]| 3D & real mass   &&& D2--brane in fluxtrap \\       
};
    \path[>=latex,->,font=\scriptsize] (m-2-2) edge node[description] {reduction} (m-5-2);
    \path[>=latex,<->,font=\scriptsize] (m-2-5) edge node[description] {T--duality} (m-5-5);
    \path[>=latex,<-,font=\scriptsize] (m-2-2) edge node[auto] {effective} (m-2-5) edge node[auto,swap] {theory} (m-2-5);;
    \path[>=latex,<-,font=\scriptsize] (m-5-2) edge node[auto] {effective} (m-5-5) edge node[auto,swap] {theory} (m-5-5);

  \end{tikzpicture}
  
  \caption{Gauge and string theory interpretation of the real mass in three and four dimensions. The reduction in gauge theory is realized as a T--duality in string theory.}
  \label{fig:strategy}
\end{figure}

Our strategy is as follows. The $(2+1)$--dimensional gauge theory with
\emph{real mass} terms that we consider can be understood as coming
from the reduction of $(3+1)$--dimensional gauge theories with Wilson
line boundary conditions for a global symmetry. This boundary condition
in turn has a
natural string theory interpretation in terms of D3--branes embedded
in flat space with discrete identifications. Such backgrounds have
been rediscovered a number of times, starting from the work of
Melvin~\cite{Melvin:1963qx}, and have taken different names, such as
\emph{fluxbranes} or $\Omega$--\emph{deformed} flat space.  Since the
string theory realization of the reduction from 3+1 to 2+1
dimensions can be achieved via a T--duality in a direction parallel to
the D3--brane, we can give a string theory construction
of the real mass in terms of D2--branes living in the T--dual of the 
fluxbrane background, that we will refer to as a \emph{fluxtrap}.  The different
interpretations are summarized in Figure~\ref{fig:strategy}. The setup
we will be using is summarized in Table~\ref{tab:D2-embedding}.

\begin{table}
  \centering
  \begin{tabular}{lcccccccccc}
    \toprule
    direction & 0 & 1 & 2 & 3 & 4 & 5 & 6 & 7 & 8 & 9 \\  \midrule 
    fluxtrap & $\times $ & $\times $ & $\times $ & $\times $ & & & & & & $\times$ \\
    D2 & $\times $ & $\times $ & $\times $ \\ \bottomrule
  \end{tabular}
  \caption{Embedding of the D2--brane with respect to the
    fluxtrap. The $\times $ indicates a direction parallel to the
    brane (for the D2) or in which the bulk geometry is flat (for the fluxtrap).}
  \label{tab:D2-embedding}
\end{table}

The fluxtrap background described in this paper serves to give
twisted masses to the chiral multiplets in a brane
construction realizing the two-dimensional gauge theories in the 
gauge/Bethe correspondence of Nekrasov and Shatashvili~\cite{Nekrasov:2009uh,Nekrasov:2009ui}.  The full construction including NS5--branes and D4--branes in the fluxtrap background
will be discussed briefly here, leaving more detailed elaboration to 
future work.\footnote{In an earlier paper~\cite{Orlando:2010uu}, a brane
construction was discussed based on the Hanany--Hori type of configuration~\cite{Hanany:1997vm}, which reproduced certain aspects of the gauge theories but omitted the twisted masses.
That construction differs in certain important ways from the one of relevance here, which
reproduces all terms in the action of 
the gauge theories of~\cite{Nekrasov:2009uh,Nekrasov:2009ui} precisely,
twisted masses included.}

The plan of this note in as follows. In Section~\ref{sec:realmass}, we give a detailed
introduction to the
real mass deformations of $(2+1)$--dimensional gauge theory, with
emphasis on points useful to our construction in particular.

In Section~\ref{sec:fluxtrap}, we introduce the fluxbrane solution, which
is equivalent to turning on an Omega background, and its T--dual  (fluxtrap) that will give real
masses to the adjoint fields living on a stack of D2--branes.  We give the
explicit supergravity solution in Section~\ref{sec:bulkfields}.  
The Killing spinors preserved by
the fluxtrap solution are given in
Section~\ref{sec:bulk-supersymmetry}, and
the relationship to 
the $\O$-deformation of 4D gauge theory is discussed in Section~\ref{sec:fluxtrap-solution}.

 In
Section~\ref{sec:D2-in-fluxtrap}, we describe the three dimensional
gauge theory living on the worldvolume of D2--branes extended in the
$(x_1,x_2)$ directions. The action, fermionic symmetries and preserved
Killing spinors are detailed in Section~\ref{sec:acsymm}.  In
Section~\ref{sec:bps-excitations}, the \textsc{bps} condition is
derived and it is shown that the real mass parameter stemming from the
fluxbrane background gives exactly the physical mass of the \textsc{bps}
states. In Section~\ref{sec:low-energy-effective}, the low energy
effective action is derived to quadratic order in the fields, where
the real mass terms appear as expected.  In Section~\ref{sec:gaugebethe}, the relation 
of our construction to the gauge/Bethe correspondence of Nekrasov and Shatashvili is discussed.
Section~\ref{sec:conclusions}
gives the conclusions. In Appendix~\ref{sec:ns5-fluxtrap} we show how
to incorporate a set of parallel of NS5 branes into the fluxtrap solution. Finally, in
Appendix~\ref{sec:conventions}, some notations and conventions are
collected.

\section{Supersymmetric real masses in field theory and string theory }\label{sec:realmass}

  In this section we will give
an exposition of the supersymmetric real mass terms for chiral
superfields in $(2+1)$ and $(1+1)$--dimensional gauge theories.   The
$(2+1)$--dimensional real mass terms were 
first written down and used in~\cite{deBoer:1997ka}, and their formal
properties were developed
systematically in~\cite{deBoer:1997kr,Aharony:1997bx}.

 The real mass term is a mass term that cannot
be thought of as a superpotential term, but comes rather
from a deformation of the \textsc{susy} algebra itself.
A real mass implies a deformation of the \textsc{susy} algebra and
also vice versa: the real mass term is not supersymmetric under
the undeformed \textsc{susy} algebra, and the deformation of the algebra automatically
imparts a mass to the fields with non-vanishing eigenvalues under the central charge $Z$
by which the algebra is deformed.  This connection gives a clue to the construction
of the real mass in string theory.  In order to make our
string theory embedding maximally clear, we will give a
construction of real mass terms in a physical and superspace-free
form, that can be thought of as involving a lift to one higher
dimension.

The ``twisted masses'' in $1+1 $ dimensions in the sense of~\cite{Hanany:1997vm} and the ``real masses'' in 2+1 dimensions~\cite{deBoer:1997ka,deBoer:1997kr,Aharony:1997bx} are related.  The principle of their relationship is that ``twisted'' mass terms in $1+1$ dimensions descend from ``real'' mass in $(2+1)$ dimensions upon dimensional reduction on a circle. 
Even though it is in general possible to switch on a second, imaginary component for the twisted mass term, in the system of~\cite{Nekrasov:2009uh,Nekrasov:2009ui}, only one real component of the complex mass term is ever activated, and that real component can be taken to be the one corresponding to a local deformation in $2+1$ dimensions. With this restriction of the twisted mass parameter to a real value, the real mass term in $2+1 $ dimensions and twisted mass term in $1+1 $ dimensions correspond canonically, and we will not distinguish them.

\subsection{Real mass deformation of the D2--brane theory}
\label{sec:realmassbrane}

 For most of this paper we will focus
on a real-mass deformation of the string theory embedding of maximally supersymmetric
Yang--Mills theory in 2+1 dimensions, which corresponds to a D2--brane in flat space.
  Our treatment will closely
follow that of~\cite{Aharony:1997bx}, though we will keep our discussion independent
of superspace, and we will emphasize the interpretation of the real mass term via
dimensional oxidation on a circle. 

A D2--brane in flat space preserves 16 supercharges among the 32
of Type IIA string theory.  This is evident -- the ambient space is
flat and there is only one type of D--brane, which is a \textsc{bps} state in
flat space~\cite{Polchinski:1995mt}.
This configuration is invariant under $SO(2,1)$
rotations in
the $012$ directions and
$SO(7) $ rotations in the $3456789$ directions. These are inherited as
the Lorentz invariance and as the global
R--symmetry group $SO(7)_{3456789}$ of the D2--brane theory.  This is
the theory that at low energies flows to the \textsc{abjm} model~\cite{Aharony:2008ug},
but the strong
coupling behavior however is not relevant for the present article -- we will always work
in the limit $g\ll s \to 0$ of weak three-dimensional gauge coupling.

The D2--brane theory has many mass deformations.  For instance there
are superpotential terms as well as FI parameters one can add while preserving
some amount of supersymmetry.  These are a
different type of mass terms than the twisted masses we will consider, in that
the FI terms and superpotential terms leave the \textsc{susy} algebra undeformed.  The twisted
mass in 1+1 dimensions and the real mass in 2+1 dimensions have the
property that their introduction always requires a (central)
deformation of the \textsc{susy} algebra itself.  This is mentioned
in~\cite{Hanany:1997vm} for twisted masses in 1+1 dimensions and
in~\cite{deBoer:1997ka,deBoer:1997kr,Aharony:1997bx} for real masses in 2+1 dimensions.
The presence of the central extension is a logical
necessity -- the mass term
amounts to a half-superspace integral of an integrand that would not be
invariant under the other two supersymmetries if the \textsc{susy} representations
of the chiral multiplets were undeformed.  The deformation of
the \textsc{susy} algebra by a central charge carried by the chiral multiplets
deforms the \textsc{susy} representations of the chiral multiplets and allows
half-superspace terms whose integrand could not otherwise be invariant
under the complementary half of the \textsc{susy}.

How should we think about real masses in 3D?  A real mass deformation is
defined by a particular set of ingredients:
\begin{itemize}
\item{A set of continuous
Abelian global symmetries $U(1)\ll i$, whose Hermitean generators are $q\ll i$.
These symmetries should {\itg not} be $R$-symmetries -- they should
all act trivially on the supercharges themselves. }
\item{More generally, if there is extended supersymmetry beyond ${\cal N} = 2$, 
the abelian symmetries should leave invariant at least one complex supercharge
$Q\ll\a\neq Q\dag\ll\a$.}
\item{A choice of real mass parameters $m\uu i$, one corresponding to
each of the Abelian global symmetries. }
\end{itemize}
If the symmetry generators $\hat{q}\ll i$ are indeed exact
symmetries of the dynamics, then the mass parameters $m\uu i$ can
be chosen arbitrarily.  It may at times be useful to consider an enlarged set of
approximate
symmetries $U(1)\ll i$ that are not exact symmetries of the dynamics but are broken
by specific terms, \emph{e.g.} by the superpotential.
If only some linear combinations of the
$\hat{q}\ll i$ are exactly preserved, then this imposes a consistency condition
relating the superpotential to the $m\uu i$: they must be chosen such
that a certain linear combination $Z\equiv m\uu i \hat{q}\ll i$ 
of $U(1)$ symmetries    (summation over $i$ is implied)
leaves the superpotential invariant.   This $Z$ is
identical with the central term that deforms of the \textsc{susy} algebra.

The invariance of the action, including the superpotential, under $Z$ is a necessary and
sufficient condition for consistently combining superpotential terms with a
twisted mass deformation.  In the gauge theories of Nekrasov and Shatashvili, which we consider in Section \ref{sec:gaugebethe}, this principle constrains the matter in the fundamental
and antifundamental representations to carry exactly $-\hh$ the $Z$--charge
of the matter in the adjoint representation, in order to accommodate the superpotential
\begin{equation}
  \label{superpot44} 
  W = \wt{{\cal Q}} \cc \phi \cc {\cal Q}\ ,  
\end{equation}
where ${\cal Q}, \phi$ and $\wt{{\cal Q}}$ are the fundamental, adjoint and
antifundamental chiral multiplets, respectively.

Given these two ingredients -- the symmetries $\hat{q}\ll i$ and
the mass parameters $m\uu i$ satisfying the consistency condition --
we can define a ``real mass'' deformation for a
$(2+1) $--dimensional \textsc{susy} theory with ${\cal N} = 2$ \textsc{susy}, in a canonical way.  To describe the deformation, introduce a set of spurious,
non-dynamical ${\cal N} = 2$ Abelian vector multiplets, one for
each of the global symmetries $U(1)\ll i$.  Then minimally couple
these non-dynamical vector multiplets to the rest of the theory as
dictated by gauge invariance and ${\cal N} = 2$ supersymmetry.  Note
that the ``complex masses'' ({\itg i.e.} the quadratic terms in the
superpotential) need not vanish in order
for the supersymmetric minimal coupling to be well defined, nor even for
the quadratic terms nor the superpotential as a whole
to respect
the symmetries $U(1)\uu i$ separately.
 All that is
needed is for the superpotential (and the rest of the action) to be invariant
under the 
combination $Z = m\uu i \hat{q}\ll i$, which is a weaker condition.

Now we give the prescription for defining the full deformation of the action.
In three dimensions, an ${\cal N} = 2$ vector multiplet
contains a single real scalar $\s = \s\uu *$,
as well as a gauge field
and a Dirac fermion.  Let them be normalized such that the kinetic term
for the gauge field would be 
\begin{equation}
  \label{NondynamicalGaugeFieldNornalization}
  - {1\over{4\cc g\ll 3\sqd}} F\ll{\m\n} F\uu{\m\n}
\end{equation}
and the kinetic term for the scalar would be
\begin{equation}
  \label{NondynamicalScalarNormalization}
  - {1\over{2\cc g\ll 3\sqd}}(\pp\ll\m \s)(\pp\uu\m \s)
\end{equation}
With these
normalizations, the gauge field and the scalar have mass dimension $1$,
and the \textsc{susy} transformations are coupling independent.
Keeping only the space- and time-independent vevs $\braket{\s\uu i} $
of the real scalars $\s\uu i$ and setting them
equal to the mass parameters,
\begin{equation}
  \braket{\sigma^i} = m\uu i \ ,   
\end{equation}
we obtain a deformation of the action for the dynamical
degrees of freedom.

Define the normalizations of the charges $q\ll i$
to be coupling-independent.  That is to say, under a constant gauge
transformation $\chi = \theta = {\rm const}.$, normalized such that
$\theta = 2\pi$ is the smallest nonzero value of $\theta$ that defines
a trivial gauge transformation,  each chiral multiplet with
charges gets a phase of $\exp[i q\ll i \theta\uu i]$.  Then the real
mass terms are such that each chiral multiplet with charges $q\ll i$
gets a mass  $q\ll i m\uu i$.

In order for this to be consistent, the \textsc{susy} algebra must be deformed
by a real central charge $Z = m\uu i q\ll i$.  Suppose the undeformed
\textsc{susy} algebra is
\begin{equation}
 \{  Q\ll a , Q\dag\ll b   \} =
- 2 ( \G\ll\m\G\ll 0 )\ll{ a b} \cc P\uu\m\ ,
\end{equation}
where we use the standard sign convention $ 0 < H = + P\uu 0 = - P\ll 0 $,
and $\m$ runs from $0$ to $2$.  Then when
the masses are turned on, the
central charge $Z$, normalized as defined above, enters as
\begin{equation}
  \{ Q\ll a , Q\dag\ll b  \} = - 2 ( \G\ll\m\G\ll 0 )\ll{ a b} \cc
P\uu\m - 2 i \cc Z\cc (\G\ll 0)\ll{a b}\ .    
\end{equation}

This description of the real mass deformation is
completely equivalent to the description in~\cite{Aharony:1997bx,Hanany:1997vm}.
The exact same construction applies to construct ``twisted mass''
deformations in 2+1 dimensions, the difference being that the vev of
the spurious vector multiplet scalars $\s\uu i$ are now complex
$\s\neq \s\dag$, and so the twisted mass parameters $m\uu i$ can be
complex instead of real.  

\subsection{Lift to ${\cal N} = 1$ theories in 3+1 dimensions on a circle}

For $(2+1)$--dimensional ${\cal N} = 2$
theories that lift to $(3+1)$--dimensional ${\cal N} = 1$
theories by dimensional oxidation on a circle, there 
is a simpler way of understanding the real mass deformation,
including the normalizations.  For the construction of the
real mass deformation to lift correctly, it's important that the
$U(1)$ symmetries $\hat{q}\ll i$ that enter the central charge $Z$
should lift to exact symmetries in four dimensions, rather than
just emerging as accidental symmetries upon compactification to
2+1 dimensions and integrating out of Kaluza--Klein modes.

Generic ${\cal N} = 2$
theories in 2+1 dimensions do not have a lift to four dimensions, but many
do, including the theories of present interest to us, namely maximally
supersymmetric gauge theory.   From the point of view
of string theory, this dimensional oxidation to 3+1 dimensions is
a T--duality on a coordinate $x\uu 8$ transverse to the D2--brane,
to a T--dual coordinate $\wt{x}\uu 8$
longitudinal to the a D3--brane.  The size
of the coordinate is of course fixed by consistency of the relation
between gauge couplings.  If the radius of the circle of compactification
is 
$\wt{R}$, and the four dimensional gauge coupling is $g\ll 4$, then
the relationship is
\begin{equation}
  \frac{2\pi \wt{R}}{g\ll 4\sqd} = \frac{1}{g\ll 3\sqd}\ .  
\end{equation}

How do we think of the real mass deformation in 4-dimensional terms?
We simply lift the three-dimensional ${\cal N} = 2$ vector multiplet
in the obvious way to a four-dimensional ${\cal N}=1$ vector
multiplet, which contains a gauge field $A\ll{0,1,2,\eit}$, a Weyl
gluino, and no scalars.  With the normalizations in
equation~\eqref{NondynamicalScalarNormalization}~and~\eqref{NondynamicalGaugeFieldNornalization}
for the real scalar $\s$ and gauge field in the non-dynamical
three-dimensional vector multiplet, the field $\s$
is identified with the zero mode piece of the $x\uu \eit$ component
$A\ll \eit$ of the non-dynamical four-dimensional gauge field, with unit coefficient:
taking $A\ll \eit$ to be constant in the
$x\uu \eit$ direction, then
\begin{equation}
  \s = 1\cdot A\ll \eit \ .    
\end{equation}
The coefficient of proportionality
can be determined from the relative normalizations of the kinetic
terms for the (spurious, non-dynamical) fields
$\s$ and $A\ll \m$ that we have coupled minimally
to the $(2+1)$--dimensional theory.
So when we set the fictional vector multiplet scalar $\s\uu i$ equal to
the corresponding mass parameter $m\uu i$, this is the same
thing as setting $A\ll \eit\uu i$ to $m\uu i$ in the four-dimensional
lift.  In other words, this is a compactification with a Wilson line
boundary condition for fields charged under the symmetries $U(1)\ll i$
such that, when parallel transported around the circle, every field
transforms to itself up to the action of the monodromy
\begin{equation}
  \hat{U}\ll 8 \equiv \exp [i \sum\ll i
    \alpha\uu i \hat{g}\ll i] \ ,
\end{equation}
where
\begin{equation}
  \alpha\uu i \equiv \oint \ll 0 \uu{2\pi \wt{R}}  \di x\uu \eit \cc A\ll \eit\uu i
  = 2\pi \wt{R}\, \s\uu i\ .
\end{equation}
We are setting $\s\uu i$ to $m\uu i$, which means
\begin{equation}
  \alpha\uu i = 2\pi \wt{R}\, m\uu i\ .
\end{equation}
Therefore in the cases where the three-dimensional theory lifts to
four dimensions (with the appropriate symmetries intact), the real mass term in the three-dimensional theory can be
obtained by starting with
the undeformed four-dimensional theory and compactifying down to three
dimensions on a circle of radius $\wt{R}$ with monodromy
\newcommand{\expc}[1]{{\rm exp}\left[ {#1} \right] }
\begin{equation}
  \label{u3derivation}
  \hat{U}\ll 8
  \equiv \expc{2\pi i \wt{R} \cc \sum\ll i m\uu i \hat{g}\ll i}
\end{equation}
around the $x\uu \eit$ direction, in the limit where $\wt{R}\to 0$.

In this language the consistency conditions for twisted mass deformations
are particularly transparent.  It's clear that
one can pick any symmetries $\hat{g}\ll i$ and mass parameters
$m\uu i$ that one likes, as long as the Wilson line
compactification preserves at least ${\cal N} = 2$ \textsc{susy} in 2+1 dimensions,
the criterion for which is that $Z \equiv \sum m\ll i \hat{q}\uu i$
is a non-R symmetry with respect to at least one four-dimensional
Weyl doublet of supercharges $Q\ll \a$.  So if the four dimensional
theory has only ${\cal N} = 1$ and no extended supersymmetry, then this
just means the combination $Z$ must be a non-R global symmetry.  If
there is extended \textsc{susy} in four dimensions, then the condition is that
the action of $Z$ on supercharges must have at least one element in its
kernel.

In the case of interest, the four dimensional theory is ${\cal N} = 4$
super-Yang--Mills in 3+1 dimensions.  Its only continuous global
symmetries are the $SO(6)\simeq SU(4)$ R--symmetry group.  This
group acts on $(3+1)$--dimensional Weyl supercharges $Q\ll \a\uu A$ in the
fundamental representation ${\bf 4}$ of $SU(4)$.  We are only interested
in symmetries that preserve at least one of the four Weyl doublets,
say the fourth one $A=4$.  Then we will restrict the generators
$\hat{g}$ to lie in an $SU(3)$ subgroup that acts nontrivially on
the first three elements of the ${\bf 4}$ only.
So our Wilson line compactification is defined by some
mass parameters $m\uu i$, one for each appropriately normalized
generator $\hat{g}\ll i$ of the $SU(3) \subset SO(6)$ inside the
R--symmetry group of ${\cal N} = 4$ of super-Yang--Mills theory in
3+1 dimensions.  Then each massless four-dimensional
chiral multiplet with eigenvalues
$q\ll i$ under the generators $\hat{g}\ll i$ gets a mass in three dimensions
that is equal to $\abs{Z} = \abs{\sum q\ll i m\uu i}$.  For chiral multiplets
that are not massless in four dimensions -- if for instance they have
D--term or F--term masses given by $M\ll 4$ in four dimensions
-- then the construction makes it quite clear what their masses must be in
three dimensions, since $Z$ is a generalized momentum in the
$x\uu 8$-direction.  The mass formula at
tree-level is
\begin{align}
  \label{massformula}
  M\ll 3 &= \sqrt{M\ll 4\sqd + Z\sqd}\ , &
  Z &\equiv \sum\ll i m\uu i q\ll i \ .
\end{align}
The mass $M\ll 4$ is the same as the three-dimensional mass that comes
from F--term and D--term potentials.  So in strictly three-dimensional
terms we can write
\begin{align}
  M\ll{\text{full, tree-level}} &= \sqrt{M\ll {F + D}\sqd + Z\sqd}\ ,
  &
  Z &\equiv  \sum\ll i m\uu i q\ll i \ .
\end{align}
The full mass was computed from a classical four-dimensional dispersion relation
so of course it will be modified by perturbative quantum corrections in general
when $M\ll 4 = M\ll{F+D}$ is nonzero.  However when $M\ll 4 = M\ll{F+D}$
vanishes, then the state is massless from the four-dimensional perspective
and \textsc{bps} from a three-dimensional perspective, and the quantum corrections
to its mass should be under control -- vanishing perturbatively
and perhaps calculable nonperturbatively, as in~\cite{Hanany:1997vm}.

So the data specifying a real mass in 2+1 dimensions in the 16-supercharge
D2--brane theory are clear -- for each $SU(3)$ generator $\hat{g}\ll a$
pick a parameter $m\uu a$, and the real masses in the three-dimensional
sense are equal to eigenvalues of the operator $Z \equiv m\uu a
\hat{g}\ll a$ acting on chiral multiplets. The deformed theory can be thought of as coming from the
compactification on the D3--brane theory on a circle of radius
$\wt{R}$, with a monodromy given by $\hat{U}\ll 8 \equiv
\exp[ 2\pi i \wt{R} Z ]$.
Having noticed that the real mass can be realized by
dimensional reduction with monodromy, let us use that description to
find a string embedding of the D2--brane theory with a twisted mass.

\subsection{String embedding of the twisted mass for ${\cal N}=8$ SYM in
$D=3$.}

Consider an isolated D2--brane (we could equally well consider a
set of $N$ D2--branes)
whose gauge coupling is $g\ll 3$.  We want to lift to a D3--brane
theory on a circle of radius $\wt{R}$.  The relation between gauge
couplings is simply
\begin{equation}
  \frac{2\pi \wt{R}}{g_4^2} = \frac{1}{g\ll 3\sqd} \ ,
\end{equation}
so
\begin{equation}
  g\ll 4 = g\ll 3 \cdot \sqrt{2\pi \wt{R}} \ .
\end{equation}
So now let us consider a D3--brane extended in directions $012\eit$ in
flat spacetime with line element
\begin{equation}
  \wt{ \di s}^2 = \di \vec x_{0 \dots 2}^2 + \di \wt x_8^2 + \di \vec y_{1 \dots 6}^2 \, ,
\end{equation}
where
\begin{align}
  \di \vec x_{0 \dots m}^2 &= - \di x_0^2 + \di x_1^2 + \dots + \di
  x_m^2 & \text{and} &&   \di \vec y_{1 \dots m}^2 &= \di y_1^2 + \dots +
  y_m^2\, .
\end{align}
The tilde denotes that the direction $\wt{x}\uu 8$ is going to be
the T--dual of the $x\uu 8$ direction transverse to the twobrane that
will be the object of our ultimate interest.  The threebrane in type
IIB string theory is a \textsc{bps} state that preserves sixteen
supercharges.  We wish to compactify the $\wt{x}\uu 8$ direction
with radius $\wt{R}$.  However a straightforward identification
$\wt{x}\uu 8 \simeq \wt{x}\uu 8 + 2\pi \wt{R}$ would leave
all sixteen supercharges unbroken and would not generate a mass term.
It also would impose periodic boundary conditions on the fields living
on the D3--brane, whereas we want to impose boundary conditions
twisted by the monodromy $\hat{U}\ll 8$ given in
equation~\eqref{u3derivation}.

The only consistent way to do that in string theory is just to impose that
same monodromy on the compactification of spacetime as a whole.  From the
point of view of the spacetime as a whole, the $SO(6)$ generators of
the D3--brane gauge theory are rotations of the six directions transverse
to the threebrane, which in this case are $y_1, \dots, y_6 $.
  So the $SO(6)$ of the gauge theory just acts on the
coordinates $y_i$ in the vector representation in an obvious way.
We are interested in preserving at least ${\cal N} = 2$ supersymmetry
in three dimensions, which forces us to restrict ourselves to
an $SU(3)$ subgroup of $SO(6)$, which imposes a choice of complex structure
on $y_i$--space.  To focus on the more
constrained case of ${\cal N} = 4$ supersymmetry in three dimensions, we can
restrict the rotation to an $SU(2)$ subgroup, in which case there are a
triplet of such complex structures, but we will just focus on one for
simplicity.

Either way, we choose a complex structure on $y_i$--space.  So define
\begin{align}
  w_1 &\equiv y_1 + \imath y_2 \, ,  &
  w_2 &\equiv y_3 + \imath y_4 \, , &
  w_3 &\equiv y_5 + \imath y_6 \, ,
\end{align}
then the condition to preserve at least ${\cal N} = 2$ \textsc{susy} in 3D is
that the generators $\hat{g}$ of the monodromy lie in a subgroup
that acts as traceless Hermitean matrices on the
three complex coordinates $w_p$.  The condition to preserve
${\cal N} = 4$ in 3D is that the Hermitean matrices
$\hat{g}$ additionally lie in an $SU(2)$ subgroup -- that is,
they have a common zero eigenvalue.
It is now to that most supersymmetric case to which we would like to turn our attention.  

We take the rotation matrices to lie in the $SU(2)$ subgroup
that acts only on the directions $w_{1,2}$, and leaves
$w_3$ alone.  Since we are compactifying only one dimension,
we have only one linear combination of generators to worry about,
so we pick $m\uu a g\ll a$ to be $m\cc \s\ll 3$, where $\s\ll 3$ is
the Pauli matrix acting on the directions $w_{1,2}$ and leaving
$w\ll 3$ invariant.

According to our prescription, we should impose
$\hat{U}\ll 8 = \expc{2\pi i m\cc \wt{R} \cc \s\uu 3}$ as a monodromy
around the $\wt{x}\uu 8$ direction, which we compactify with radius
$\wt{R}$.  This is equivalent to identifying the flat, ten-dimensional
space by the combined identification (as
opposed to two independent identifications) as follows:
\begin{align}
  \label{OmegaIdents}
  \wt{x}\uu 8 &\simeq \wt{x}\uu 8 + 2\pi \wt{R} \, ,
  & \begin{pmatrix} w\ll 1 \\ w\ll 2 \end{pmatrix} &\simeq \hat{U}\ll
  8 \cdot \begin{pmatrix} w\ll 1 \\ w\ll 2 \end{pmatrix} \ .
\end{align}

This space, with these identifications, defines a purely closed-string
background in which the D3--brane can be thought of as a probe.  So for
now, let us focus on the description of the closed string background
itself.

\section{Closed string fluxbrane and fluxtrap backgrounds} 
\label{sec:fluxtrap}

This type of space, obtained by taking a quotient by identifications
of the form~\eqref{OmegaIdents},
 has been studied already very well.  Spaces of this kind, defined by a simultaneous
identification by a shift of one direction and a rotation of
some other directions, go by the name of ``fluxbranes'' and have
been studied for quite some time, starting with work in
the pure general relativity context by
Melvin~\cite{Melvin:1963qx}. 
The ``flux'' in ``fluxbrane'' refers to the idea of starting in 5D
general relativity with one circle compactified \emph{\`a la}
Kaluza--Klein.

In the case of an $S\uu 1$ compactification with a monodromy $\hat{U}$
around the $S\uu 1$ acting on some other space ${\bf X}$, it is
natural to think of ${\bf X}$ as fibered over the circle, with the
circle as the base and the fibration data defined as gluing maps on
the fiber, specified by the monodromy.  But the space carries a \it
second \rm fibration structure in which the $S\uu 1$ is the fiber
and {\bf X} the base of the fibration. 

In the picture where the $S\uu 1$ shift-circle is the fiber direction,
the fibration structure is nontrivial even locally, in the sense of there being a
local curvature of the connection of the bundle.  In other words,
there is Kaluza--Klein flux.  This is not true in the original picture,
where the $S\uu 1$ is the base and the fibration of ${\bf X}$ over it
is described by a connection that is locally flat since the base is
one-dimensional.
The ``fluxbrane'' picture -- in which the space ${\bf X} $ is the
base and the shift-circle $S\uu 1$ the fiber -- is the more natural
in one in the Kaluza--Klein theory of 4D general relativity and
electromagnetism, or any theory in which the $S\uu 1$ is
taken to be small.  That is the picture in which these spaces
are thought of in \emph{e.g.} ~\cite{Melvin:1963qx,Tseytlin:1994ei,Tseytlin:1995zv,Russo:2001na,Gutperle:2001mb,Takayanagi:2001jj,Hellerman:2006tx}, whence the name ``fluxbrane''.
For us, the shift-circle $S\uu 1$ is the direction $\wt{x}\uu 8$
in the type IIB string theory, and the Euclidean directions
$w\ll {1,2}$ are the space ${\bf X}$, and indeed when we think of
${\bf X} $ as the base, there really is Kaluza--Klein flux,
as we shall now see.

\subsection{Bulk fields and T--duality transformations}
\label{sec:bulkfields}
In the following, we specify the fluxbrane background in cylindrical
coordinates. We will then perform a T--duality along the
direction $x_{\wt{8}}$ and derive the expressions for the metric, vielbein,
$B$--field and the dilaton for the fluxtrap. This resulting geometry
will provide the closed string background for the D2--branes in the
following sections.

\paragraph{Fluxbrane.}

In cylindrical coordinates, defined by
\begin{align}
  \rho_1 \eu^{\imath \theta_1} &\equiv w_1 = y_1 + \imath y_2 \, &
  \rho_2 \eu^{\imath \theta_2} &\equiv w_2 = y_3 + \imath y_4 \, & 
  x_3 + \imath x_9 &\equiv y_5 + \imath y_6 \, , & 
  \wt x_8 = \wt R \wt u \, ,
\end{align}
our fields have the following simple form:
\begin{align}
  \wt g_{\mu \nu} \di \wt X^\mu \di \wt X^\nu &= \di \vec x_{0 \dots 3}^2 + \di
  \rho_1^2 + \rho_1^2 \di \theta_1^2 + \di \rho_2^2 + \rho_2^2 \di
  \theta_2^2 + \wt R^2 \di \wt u^2 + \di x_9^2  \, , \\
  \wt B_{\mu \nu} \di \wt X^\mu \di \wt X^\nu &= 0 \, , \\
  \wt \Phi &= \log ( \wt R\, g_3^2 ) \, , \label{eq:phi}
\end{align}
where $\wt X^\mu = ( x_0,\dots , x_3, \rho_1, \theta_1, \rho_2,
\theta_2, \wt u, x_9)$. The reason for our choice of the constant value~(\ref{eq:phi}) for the
dilaton~$\wt \Phi $ will become clear later
on: $g_3$ will be the gauge coupling for the effective quantum field
theory living on D2--branes at the origin.

The space $\setR^5 / \Gamma $ is obtained by imposing the 
identifications in Equation~\eqref{OmegaIdents}:
\begin{equation}
  \label{eq:theta-periodicity}
  \begin{cases}
    \wt u \simeq \wt u + 2 \pi \, k_1 \, , \\
    \theta_1 \simeq \theta_1 + 2 \pi m \wt R \, k_1 \, , \\
    \theta_2 \simeq \theta_2 - 2 \pi m \wt R \, k_1 \, ,
  \end{cases} \hspace{2em} k_1 \in \setZ \, ,
\end{equation}
in addition to the standard identifications for cylindrical coordinates,
\begin{align}
  \theta_1 &\simeq \theta_1 + 2 \pi \, k_2 \, , & \theta_2 &\simeq
  \theta_2 + 2 \pi \, k_3 \, , \hspace{2em} k_2, k_3 \in \setZ \, .
\end{align}

It is convenient to disentangle the periodicities.  For this reason we
introduce the new angular variables
\begin{equation}
  \begin{cases}
    \phi_1 = \theta_1 - m \wt R \wt u \, ,\\
    \phi_2 = \theta_2 + m \wt R \wt u  \, ,
  \end{cases}
\end{equation}
to rewrite the metric in the form
\begin{multline}
  \wt {\di s^2} = \di \vec x_{0 \dots
    3}^2 + \di \rho_1^2 + \rho_1^2 \di \phi_1^2 + \di \rho_2^2 +
  \rho_2^2  \di \phi_2^2 \\
  + 2 m \wt R \left( \rho_1^2 \di \phi_1 - \rho_2^2 \di \phi_2 \right)
  \di \wt u + \wt R^2 \left( 1 + m^2 \left( \rho_1^2 + \rho_2^2
    \right) \right) \di \wt u^2 + \di x_9^2 \, ,
\end{multline}
with the three \emph{independent} sets of identifications:
\begin{equation}
  \label{eq:identifications}
  (\wt u , \phi_1 , \phi_2) \mapsto (\wt u + 2 \pi \, n_1 , \phi_1 +
  2\pi \, n_2 ,
  \phi_2 + 2 \pi \, n_3) \, , \hspace{2em} n_1, n_2, n_3 \in \setZ \, .
\end{equation}
The space is of course still locally flat, but in this coordinate
system one can see immediately the $S^1 $ fibration structure where
the fiber is described by $\wt u$. This can be interpreted in terms of
a non-flat Kaluza--Klein gauge connection  $ \left( \rho_1^2 \di \phi_1 - \rho_2^2 \di
  \phi_2 \right)$, which explains the origin of the name
\emph{fluxbrane}. The natural vielbein is given by
\begin{equation}
  \label{eq:flat-vielbein}
  \begin{aligned}
    \wt e^n &= \di x^n \, , & n &= 0, 1, 2, 3, 9 \\
    \wt e^4 &= \di \rho_1 \, , &
    \wt e^5 &= \rho_1 \left( \di \phi_1 + m \wt R \di \wt u \right) \, , &
    \wt e^6 &= \di \rho_2 \, , &
    \wt e^7 &= \rho_2 \left( \di \phi_2 - m \wt R \di \wt u \right) \, ,\\
    \wt e^8 &= \wt R \di \wt u \, .
  \end{aligned}
\end{equation}

An alternative description for the same space can be obtained by passing
to rectilinear coordinates:
\begin{align}
 z\ll 1 \equiv  x_4 + \imath x_5 &\equiv \rho_1 \eu^{\imath \phi_1} \, ,&    z\ll 2 \equiv
 x_6 + \imath x_7
  &\equiv \rho_2 \eu^{\imath \phi_2} \, , & x_8 &\equiv \wt R \wt u \, ,
\end{align}
which are related to the previous coordinate by a rotation in $\wt u$:
\begin{align}
  \begin{pmatrix} x_4 \\ x_5 \end{pmatrix} &=
  \begin{pmatrix}
    \cos ( m \wt R \wt u) & \sin ( m \wt R \wt u) \\
    - \sin ( m \wt R \wt u) & \cos ( m \wt R \wt u)
  \end{pmatrix}
  \begin{pmatrix}
    y_1 \\ y_2
  \end{pmatrix} \, , &  
  \begin{pmatrix} x_6 \\ x_7 \end{pmatrix} &=
  \begin{pmatrix}
    \cos ( m \wt R \wt u) & \sin ( m \wt R \wt u) \\
    - \sin ( m \wt R \wt u) & \cos ( m \wt R \wt u)
  \end{pmatrix}
  \begin{pmatrix}
    y_3 \\ y_4
  \end{pmatrix} \, .
\end{align}
The metric can be recast in the form
\begin{equation}
  \label{eq:flat-metric}
  \wt g_{\mu \nu} \di \wt X^\mu \wt X^\nu = \di \vec x^2_{0 \dots 3} +
  \sum_{i = 4}^7 \left( \di x_i + m V^i \di x_8 \right)^2 + \di x_8^2 +
  \di x_9^2 \, , 
\end{equation}
where $V^i \del_i $ is the Killing vector
\begin{equation}
  \label{eq:KillVec}
  V^i \del_i = - x^5 \del_{x_4} + x^4 \del_{x_5} + x^7 \del_{x_6} -
  x^6 \del_{x_7} = \del_{\phi_1} - \del_{\phi_2} \, ,
\end{equation}
with norm
\begin{equation}
  \norm{ V }^2 = x_4^2 + x_5^2 + x_6^2 + x_7^2 =  \rho_1^2 + \rho_2^2 \, . 
\end{equation}
This is precisely the form of the $\Omega$ deformation of flat space in
the directions $z_1$ and $z_2$ with parameters $\varepsilon_1 = -
\varepsilon_2 = m$ as described
in~\cite{Nekrasov:2003rj}.

\paragraph{Fluxtrap.}

In order to make contact with the $(2+1)$--dimensional theory living on the D2--branes
 we want to describe, we now perform a T--duality in the $x_8$--direction.
Since the three identifications in Equation~\eqref{eq:identifications}
are independent we can make use of Buscher's rules~\cite{Buscher:1987sk} for the
T--duality to exchange the coordinate $\wt{u}$ for a new
coordinate $u$, also with periodicity $2\pi$. The metric and $B$-field
become:
\begin{equation}
  \label{eq:buscher}
  \begin{aligned}
    g_{\sigma\rho} &= \wt g_{\sigma\rho} + {{ \wt B_{\wt{u} \sigma}\, \wt
        B_{\wt{u}\rho} - \wt g_{\wt{u}\sigma} \, \wt
        g_{\wt{u}\rho} } \over{\wt g_{\wt{u}\wt{u}}}} \, , &
    g_{uu} &= {{(\alpha^\prime)^2}\over{\wt g_{\wt{u}\wt{u}}}}
    \, , & g_{u \sigma} &= \alpha^\prime \, {{\wt B_{\wt{u}\sigma}}\over
      {\wt g_{\wt{u}\wt{u}}}}\,, \\
    B_{\sigma\rho} &= \wt B_{\sigma\rho} + {{\wt B_{\wt{u} \sigma} \wt
        g_{\wt{u} \rho} - \wt B_{\wt{u}\rho} \wt
        g_{\wt{u}\sigma}} \over{\wt g_{\wt{u}\wt{u}}}} \ , &
    B_{u \sigma} &= \alpha^\prime\, {{\wt g_{\wt{u}\sigma}} \over {\wt
        g_{\wt{u}\wt{u}} }}\,, &
    \Phi &= \wt \Phi - \frac{1}{2} \, \log \left(
      {{\wt g_{\wt{u}\wt{u}}}\over{\alpha^\prime}} \right) \ ,
  \end{aligned}
\end{equation}
where $(\sigma,\rho)$ run over all coordinates except $\wt{u}$ or
the dual coordinate $u$. In terms of the dual radius
\begin{equation}
  R = \frac{\alpha'}{ \wt R } \, ,
\end{equation}
the new dimensionful coordinate is
\begin{equation}
  x_8 = R u \, ,
\end{equation}
such that the metric, $B$--field and dilation after T--duality are given
by
\begin{align}
  \label{eq:bulk-metric} 
  \di s^2 &= \di \vec{x}_{0 \dots 3}^2 + \di \rho_1^2 + \di \rho_2^2 +
  \rho_1^2 \di \phi_1^2 + \rho_2^2 \di \phi_2^2 +\frac{ -m^2 \left(
      \rho_1^2 \di \phi_1 - \rho_2^2 \di \phi_2 \right)^2 + \di x_8^2
  }{1 + m^2 \left( \rho_1^2 + \rho_2^2 \right)} + \di
  x_{9}^2 \, , \\
 \label{eq:b-field} 
  B &=  m\, \frac{\rho_1^2 \di \phi_1 - \rho_2^2 \di
    \phi_2}{1 + m^2 \left( \rho_1^2 + \rho_2^2 \right)} \wedge \di x_8\,, \\
\label{eq:dilaton}
  \eu^{-\Phi} &= \frac{\sqrt{1 + m^2 \left( \rho_1^2 + \rho_2^2
      \right)}}{g_3^2 \sqrt{\alpha'}} \, .
\end{align}

Observe that the complex coordinates $z_1 \equiv  x_4 + \imath x_5$
and $z\ll 2 \equiv  x_6 + \imath x_7$ are left untouched by
T--duality in the direction $\wt u$ since the three identifications in
Equation~\eqref{eq:identifications} are independent. Moreover,
$V^\mu \del_\mu$ remains a Killing vector for the geometry.

\bigskip

It is convenient to introduce a ``natural'' vielbein for the T--dual
geometry. This is obtained by imposing 
\begin{equation}
  \wt e\ud{m}{\mu}\del \wt X^\mu = {e}\ud{m}{\mu}\del X^\mu \, ,
\end{equation}
where $\del X $ is the worldsheet derivative. Under T--duality,
$\del X$ transforms as (see \emph{e.g.}~\cite{Bakas:1995hc}):
\begin{align}
  \del \wt u& \to \frac{1}{\wt g_{\wt u \wt u}}(\alpha' \del u- (\wt g_{\sigma
    \wt u}+\wt B_{\sigma \wt u}) \del X^\sigma) \, ,\\
  \del \wt X^\sigma& \to \del X^\sigma \, ,
\end{align}
where $X^\sigma$ runs again over all the coordinates other than $u$. The
invariance of $e\ud{m}{\mu}\del X^\mu$ results in
\begin{equation}
\label{eq:vielbein-T--duality}
  \begin{dcases}
    e\ud{m}{u}=\frac{\alpha'}{\wt g_{\wt u \wt u}}\wt e\ud{m}{\wt u}\,, \\
    e\ud{m}{\sigma} = \wt e\ud{m}{\sigma} -
    \frac{\wt g_{\sigma \wt u}+\wt B_{\sigma \wt u}}{\wt g_{\wt
        u \wt u}} \wt e\ud{m}{\wt u} & \text{for $X^\sigma \neq u$.}
  \end{dcases}
\end{equation}
The inverse of these transformations is given by
\begin{equation}
  \begin{dcases}
    e\du{m}{u} = \frac{\wt g_{\wt u \wt u}}{\alpha'}\wt e\du{m}{\wt u} +
    \frac{\wt g_{\sigma \wt u}+\wt B_{\sigma \wt u}}{\alpha'}
    \wt e\du{m}{\sigma}\,,   \\
    e\du{m}{\sigma} = \wt e\du{m}{\sigma}  & \text{for $X^\sigma \neq u$.}
  \end{dcases}
\end{equation}
Starting from the vielbein for flat space in
Equation~\eqref{eq:flat-vielbein}, we obtain
\begin{equation}
  \begin{aligned}
    & e^n = \di x^n \, , \hspace{2em} n = 0, 1, 2, 3, 9 \\
    &\begin{aligned}
      e^4 &= \di \rho_1 \, , &
      e^5 &= \frac{\rho_1}{\Delta^2} \left( \di \phi_1 + m^2 \rho_2^2
         \left( \di \phi_1 + \di \phi_2 \right) + m \di x_8
      \right)\, , \\
      e^6 &= \di \rho_2 \, , &
      e^7 &= \frac{\rho_2}{\Delta^2} \left( \di \phi_2 + m^2 \rho_1^2
        \left( \di \phi_1 + \di \phi_2 \right) - m
         \di x_8 \right)\, ,  
    \end{aligned} \\
    & e^8 = \frac{1}{\Delta^2} \left(  - m  \rho_1^2  \di \phi_1 + m
        \rho_2^2 \di \phi_2 +  \di x_8 \right) \, ,
  \end{aligned}
\end{equation}
where
\begin{equation}
  \Delta^2 = 1 + m^2 \left( \rho_1^2 + \rho_2^2 \right) \, .
\end{equation}   
With this, we have collected all the necessary expressions for the fluxtrap geometry.

\subsection{Supersymmetry of the closed string background}
\label{sec:bulk-supersymmetry}

After having derived the form of the metric, $B$--field, dilaton and
the vielbein in the fluxtrap background, we will now investigate the
number of supersymmetries that are preserved by this background and
explicitly give the preserved Killing spinors. It is convenient
first to study the supersymmetries preserved by the
fluxbrane background and then apply the T--duality to
transform the Killing spinors.

\bigskip
In our choice of coordinates, the Killing spinors in the \emph{flat}
background are given by
\begin{equation}
  K^{IIB} = \exp [ \tfrac{1}{2} \theta_1 \Gamma_{45} + \tfrac{1}{2}
  \theta_2  \Gamma_{67} ] \epsilon_0 \, ,   
\end{equation}
where $\epsilon_0$ is a complex Weyl spinor. Introducing $\phi_1 $ and
$\phi_2$, this becomes
\begin{equation}
  K^{IIB} = \exp[\tfrac{1}{2} \phi_1 \Gamma_{45} + \tfrac{1}{2}
  \phi_2  \Gamma_{67} ] \exp [ \frac{m \wt R \wt u}{2} \left( \Gamma_{45}
    - \Gamma_{67} \right) ] \epsilon_0 \, .
\end{equation}
Observe that all the variables are $2 \pi$--periodic.  The matrix
$\G\ll{45} - \G\ll{67}$ has eigenvalues $\pm 2i $ and $0$.  There are thus two
possibilities for the Killing spinor to have the right
periodicity~\cite{Russo:2001na}:
\begin{enumerate}
\item $m \wt R $ is an integer. In this case the original variables
  $\theta_1 $ and $\theta_2$ are only $2 \pi$-periodic and the only
  non--trivial identification in equation~\eqref{eq:theta-periodicity}
  is $\wt u \simeq \wt u + 2 \pi k_1$. In other words, the spacetime is
  the standard flat space, preserving 32~real supercharges.
\item For generic values of $m \wt R$, the second exponential is
not periodic in $\wt{u}$ unless $\G\ll{45} - \G\ll{67}$ vanishes,
in which case the dependence on $\wt u$ drops out of the spinor.
\end{enumerate}
The first case is simply flat space without any deformation; in the following we will pursue the second alternative, which cuts down the number of Killing spinors by half.  The
orthogonal projectors
\begin{equation}
  \proj{flux}_\pm = \tfrac{1}{2} \left( \Id \pm \Gamma_{4567} \right) \, ,
\end{equation}
satisfy
\begin{equation}
  \Gamma_i \proj{flux}_\pm = \begin{cases}
    \proj{flux}_\mp \Gamma_i & \text{if $i = 4,5,6,7$}, \\
    \proj{flux}_\pm \Gamma_i & \text{otherwise.}
  \end{cases}
\end{equation}
Using the fact that
\begin{equation}
  \proj{flux}_- \left( \Gamma_{45} - \Gamma_{67} \right) =
  \proj{flux}_- \left( \Id + \Gamma_{4567}  \right) \Gamma_{45} =
  \proj{flux}_- \proj{flux}_+ \Gamma_{45} = 0   \,,
\end{equation}
we find the expression for the 16 type IIB Killing spinors of the
fluxbrane background:
\begin{equation}
  K^{IIB} = \proj{flux}_- \exp [ \tfrac{1}{2} \phi_1
  \Gamma_{45} + \tfrac{1}{2} \phi_2 \Gamma_{67} ] \epsilon_0 \, .
\end{equation}

\bigskip

Having obtained the expressions for the Killing spinors in type IIB, we can
translate them into the T--dual type IIA fluxtrap picture (see
\emph{e.g.}~\cite{Bergshoeff:1994cb}): $K^{IIA} = \epsilon_L +
\epsilon_R$, where
\begin{equation}
\label{eq:trap-Killing-spinors}
\boxed{  \begin{cases}
    \epsilon_L = \eu^{-\Phi/8} \left( \Id + \Gamma_{11} \right)
    \proj{flux}_- \exp [ \tfrac{1}{2} \phi_1 \Gamma_{45} +
    \tfrac{1}{2} \phi_2 \Gamma_{67} ] \, \epsilon_0 \,,\\
    \epsilon_R = \eu^{-\Phi/8} \left( \Id - \Gamma_{11} \right)
    \Gamma_u \proj{flux}_- \exp [ \tfrac{1}{2} \phi_1 \Gamma_{45} +
    \tfrac{1}{2} \phi_2 \Gamma_{67} ] \, \epsilon_1 \,,
  \end{cases}}
\end{equation}
with $\epsilon_0 $ and $\epsilon_1 $ constant Majorana spinors, and
\begin{equation}
  \label{eq:Gamma-u}
  \Gamma_u = R\,\Delta \,  e\ud{a}{u} \Gamma_a  = \frac{m \rho_1}{\Delta} \Gamma_5 - \frac{m \rho_2}{\Delta} \Gamma_7 +
  \frac{1}{\Delta} \Gamma_8 \, 
\end{equation}
is the $\Gamma$ matrix in the $u$ direction, normalized to square to
the identity, $ \left( \Gamma_u \right)^2 = \Id$.
These spinors are such that the corresponding variations of the
dilatino and gravitino (Equations~\eqref{eq:delta-dilatino}
and~\eqref{eq:delta-gravitino}) vanish.

Both $\epsilon_0 $ and $\epsilon_1$ have 32~real components. The
projectors $\proj{flux}_-$ and $\left( \Id \pm \Gamma_{11} \right) $
each reduce the preserved supercharges by a factor of one half. In our fluxtrap background, we are
therefore left with \textbf{16~preserved real supercharges}.

\subsection{Fluxtrap solution as the string theory of the $\Omega$--background.}
\label{sec:fluxtrap-solution}

The fluxtrap solution~\eqref{eq:bulk-metric}~--~\eqref{eq:dilaton} is the string theory
of the $\O$--deformation, and we would like to understand the meaning of that.
So far we have considered branes -- specifically D2--branes -- embedded transverse
to the $z_{1,2}$ directions.  To understand the relationship with the
$\O$--deformation of four-dimensional gauge theory, let us consider a
different type of D--brane, which we shall embed to fill the $z_{1,2}$ directions
and denote with a prime.

The $\O$--deformation of maximally supersymmetric gauge
theory~\cite{Nekrasov:2003rj} can be defined by starting with
five-dimensional gauge theory and dimensionally reducing on a circle
with twisted boundary conditions defined by the identifications
\eqref{OmegaIdents}, where the rotation~\eqref{OmegaIdents} acts on
directions in the gauge theory itself, rather than on scalar fields.
That is to say, the directions of the 5D gauge theory, in our
coordinates, would be $x_{4,5,6,7,\wt{8}}$.  It is natural to
interpret this gauge theory as the dynamical theory of Euclidean
D4$\pr$--branes spanning the $x_{4,5,6,7,\wt{8}}$ directions.
 
Dimensionally reducing to 4-dimensional gauge theory, as
in~\cite{Nekrasov:2003rj} amounts to performing a T--duality along the
$\wt{x}_ 8$ direction, leaving us with the flux-trap solution~\eqref{eq:bulk-metric}~--~\eqref{eq:dilaton}.  The D4$\pr$--branes have
now been transformed into D3$\pr$--branes spanning the $z_{1,2}$
directions.  The $\O$--deformation of four-dimensional ${\cal N}=4$
super-Yang--Mills theory can be thought of precisely as the $\apr\to 0$
limit of a set of D3$\pr$--branes embedded in the fluxtrap solution~\eqref{eq:bulk-metric}~--~\eqref{eq:dilaton}, spanning the directions
$z_{1,2}$.
 
The prime on the D3$\pr$--branes emphasizes that the four-dimensional
gauge theory here is \emph{not} to be identified with the gauge
theory from which we constructed our three-dimensional theory with
twisted masses.  The two types of branes are entirely separate, and
not to be included simultaneously in the same dynamical system.
(Indeed, the primed branes live in type IIB string theory and the
unprimed branes in type IIA, although this is not significant -- a
T--duality along a trivial direction such as $x_{1,2,3,9}$ transforms
a IIA brane into a IIB brane and leaves the fluxtrap solution
unaffected.)
 
The primed D3--branes, on which the $\O$--deformed gauge theory lives,
are Euclidean and space-filling in the $z_{1,2}$ directions.  The
unprimed D2--branes, on which the gauge theory
of~\cite{Nekrasov:2009uh} lives, are transverse to the $z_{1,2}$
directions.  The $\O$--deformation appears in the former as a
position-dependent gauge coupling, and in the latter as a twisted mass
term.
 
The relationship between the twisted mass deformation of the unprimed
D2--branes and the $\O$--deformation of the primed D3--branes is that
both arise from the same deformation of the closed string background
in which each type of brane is embedded in its own way.

We expect that embedding the $\O$--deformation in string theory via
the fluxtrap solution clarifies and simplifies certain aspects of the
$\O$--deformation that otherwise appear somewhat technical and opaque.
Let us take an easy example: one particularly salient feature of the
$\O$--deformation is its localization of instantons to the origin of
the four spacetime dimensions of the gauge theory.  Even from the
gauge theory perspective, a moment's thought makes it clear that such
a localization can only come about through a position-dependent gauge
coupling with a maximum at the origin.  A small instanton is
pointlike, and cannot therefore couple to a metric or $B$--field; its
only interaction with background fields is through the gauge coupling.
Since its action is inversely proportional to $g\ll 4\uu{\prime 2}$,
the instanton's action is minimized where the gauge coupling attains
its maximum value.
 
For a D3'--brane in the flux-trap solution, the four-dimensional gauge
coupling is
 \begin{equation}
   g\ll {4\cc\O}\uu{\prime \cc 2}  = \frac{g\ll {4\cc(0)}\uu{\prime\cc 2}}{\sqrt{1 +\bar{\varepsilon}\sqd r\sqd}}\ ,    
 \end{equation}
 where
 \begin{equation}
   g\ll{4\cc{(0)}}\pr \equiv (2\pi)\uu{1/2} \cc (\apr)\uu{1/ 4}\cc g\ll 3     
 \end{equation}
 is the local gauge coupling of the four-dimensional gauge theory near the origin,
 and
 \begin{equation}
\bar{ \varepsilon}\equiv m\ .
  \end{equation}
  
  We have introduced the primed branes only to clarify the relationship between
  the mass-deformed 3D theory and the $\O$--deformed 4D Euclidean theory.  Hereafter
  we shall leave the primed branes and not return to them in the present article.  However
  we anticipate that the string theory embedding of the $\O$--deformation may be useful
  for analyzing the $\O$--deformed theory on the primed D3$\pr$--branes.

\section{Open strings}
\label{sec:D2-in-fluxtrap}

In the following section, we describe the three-dimensional
gauge theory that lives on the worldvolume of a single D2--brane extended in
the directions $x_1$ and $x_2$ in the fluxtrap background (see
Table~\ref{tab:D2-embedding}). After briefly discussing the
kappa-symmetry-fixed action in the \emph{static gauge}, we derive the
expressions for the eight Killing spinors preserved by the D2--brane
located at $\rho_1=\rho_2=0$. We then derive the supersymmetry
generators $Q$ and find a \emph{rotating brane solution} which saturates the
\textsc{bps} bound and preserves four supercharges. Finally we show
how the low energy description of the D2--brane dynamics contains the
expected form of a real mass term.  The dynamics of N identical D2--branes in
the same background can then be inferred up to commutator terms from the single-trace
form of the D--brane action, since we are working at string tree level.

\subsection{Action and fermionic symmetries}\label{sec:acsymm}

We would like to describe the dynamics of a D2--brane extended in the
$(x_1,x_2)$ directions. Since the
background is symmetric under translations in the $x_1$ and $x_2$
directions, we can choose a consistent truncation of the theory where:
\begin{itemize}
\item the two-form on the D2--brane is vanishing,
  \begin{equation}
    B_{\alpha \beta} + 2 \pi \alpha' F_{\alpha \beta} = 0 \, ;
  \end{equation}
\item the position of the D2--brane in the transverse direction only
  depends on time.
  \item{The coordinates $x\ll {3,8,9}$ are constant and the gauge field is flat.}
\end{itemize}
This truncation can be realized as the restriction to the subset of configuration space
invariant under a set of discrete symmetries and translational invariances.
Under this truncation, the relevant part of the bosonic action and
the kinetic term for the fermions~\cite{Martucci:2005rb} are in our conventions (see
Appendix~\ref{sec:conventions}):
\begin{equation}
  S = - \mu_2 \int \di^3 \zeta \, \eu^{-\Phi} \sqrt{- \det g_{\alpha
      \beta} } \left( 1 - \frac{\eu^{\Phi/4}}{2} \bar \theta \left( \Id -
      \Gamma_{D2} \right) g^{\alpha \beta} \Gamma_\beta \del_\alpha \theta
  \right) +O \left ( \th\uu 4 \right )\, ,
\end{equation}
where $\mu_2 = \left(2 \pi \right)^{-2} (\alpha')^{-3/2} $, $\theta = \theta_L + \theta_R$ is a Majorana spinor, $ g_{\alpha
  \beta}$ is the pullback of the metric on the D2--brane,
\begin{equation}
  g_{\alpha \beta} = g_{\mu \nu} \frac{\del X^\mu}{\del \zeta^\alpha
  } \frac{\del X^\nu}{\del \zeta^\beta} \hspace{2em} \alpha, \beta = 0, 1, 2 \, ,
\end{equation}
$\zeta^\alpha $ are the intrinsic coordinates on the worldvolume of the D2--brane and
$\Gamma_{D2} $ is given by~\cite{Bergshoeff:1996tu},
\begin{equation}
  \Gamma_{D2} = \frac{1}{\sqrt{-\det g_{ab}}} \frac{\epsilon^{\alpha \beta \gamma}}{3!} \Gamma_\alpha \Gamma_\beta \Gamma_\gamma \, ,  
\end{equation}
where $\Gamma_{\alpha } $ is the pullback of the gamma matrices on the
brane\footnote{The normalization factors are chosen such that
  $\Gamma_{D2}$ squares to the identity: $ ( \Gamma_{D2} )^2 = \Id $. }:
\begin{equation}
  \Gamma_\alpha = \frac{\del X^\mu}{\del \zeta^\alpha } e\ud{m}{\mu} \Gamma_m \, .
\end{equation}

Since $\del_{x^0} $, $\del_{x^1}$ and $\del_{x^2} $ are Killing
vectors for the bulk metric in Equation~\eqref{eq:bulk-metric}, it is easy to fix
reparametrization symmetry of the intrinsic coordinates $\zeta$ by
choosing a static gauge for the embedding:
\begin{align}
  x^0 &=\zeta^0 \, , & x^1 &= \zeta^1 \, , & x^2 &=\zeta^2 \, .
\end{align}
The corresponding pullback of the metric is simply
\begin{equation}
  \label{eq:D2-general-pullback-metric}
  g_{\alpha \beta} \di \zeta^\alpha \di \zeta^\beta =  \hat g_{00}( \di \zeta^0  
  )^2 + ( \di \zeta^1 )^2 + ( \di
  \zeta^2 )^2  \, ,
\end{equation}
where $\widehat g_{00} = -1 + \del_0 X^\rho \del_0 X^\sigma
g_{\rho\sigma}$ and $X^{\rho} $ and $X^{\sigma}$ run over the
transverse coordinates,
\begin{equation}
  X^\rho, X^\sigma = \set{ x_3, \rho_1, \phi_1, \rho_2, \phi_2, x_8, x_9 } \, .
\end{equation}
The fact that the $B$ field does not contribute can be understood by
observing that $\del_{x_1} $ and $\del_{x_2}$ are Killing vectors and
a double T--duality in these directions maps our D2--brane to a D0
brane. 

\bigskip

The action is invariant under kappa-symmetry and under the susy
transformations induced by the bulk Killing spinors $\epsilon$. On the
fermionic variable $\theta$ they act as follows:
\begin{align}
  \delta_\kappa \theta &= \left( \Id + \Gamma_{D2} \right) \kappa \, , &
   \delta^{\text{susy}}_\epsilon \theta &= \epsilon \, ,
\end{align}
where $\kappa$ is a Majorana spinor. The transformation
$\delta_\kappa$ can be used to impose a covariant gauge fixing,
\begin{equation}
  \Gamma_{11} \,\theta = \theta\ \ \Rightarrow\ \theta_R=0,
\end{equation}
in order to obtain the same number of bosonic and fermionic degrees of
freedom. After gauge fixing, the kinetic term of the fermionic
action takes the form
\begin{equation}
  S_f = \frac{\mu_2}{2} \int \di^3 \zeta \, \eu^{-3\Phi/4} \sqrt{-\det
    g_{\alpha\beta}} \, \bar\psi \, g^{\alpha\beta} \Gamma_\beta
  \del_\alpha \psi \, ,
\end{equation}
and using the form of the pullback in
Equation~\eqref{eq:D2-general-pullback-metric}:
\begin{equation}
  S_f = - \frac{\mu_2}{2} \int \di^3 \zeta \, \frac{\eu^{-3\Phi/4}}{ \sqrt{-\det
    g_{\alpha\beta}} } \, \bar\psi \,  \hat \Gamma_0 \dot \psi \, ,
\end{equation}
where $\psi $ is the Majorana--Weyl spinor $\psi = \theta_L$ and $\hat
\Gamma_0$ is the pullback of the gamma matrices in the direction $\zeta^0$:
\begin{equation}
  \label{eq:Gamma0-hat}
  \hat \Gamma_0 = \left. \Gamma_\alpha  \right|_{\alpha = 0} = \frac{\del X^\sigma}{\del \zeta^0} e\ud{m}{\sigma} \Gamma_m \, .  
\end{equation}
The action is invariant under the transformation
\begin{equation}
  \label{eq:D2-fermionic-transformations}
  \delta_\epsilon \psi =  \left( \delta^{\text{susy}}_\epsilon  -
  \left. \delta_\kappa\right|_{\kappa=\epsilon_R} \right) \psi = \epsilon_L - 
  \Gamma_{D2} \, \epsilon_R \, ,
\end{equation}
which leaves $\theta_R $ invariant, consistently with the gauge choice
$\theta_R=0$.

\paragraph{Supersymmetries preserved by the static embedding.}

We say that a Killing spinor $\epsilon = \epsilon_L + \epsilon_R$
is \emph{preserved} by the D2--brane if the associated transformation leaves
$\psi $ invariant:
\begin{equation}
  \delta_{\epsilon} \psi = \epsilon_L - \Gamma_{D2} \, \epsilon_R = 0 \, .
\end{equation}
If we choose the \emph{static embedding} in which $\hat
g_{00} = 1$, the expression of $\Gamma_{D2}$ is simply
\begin{equation}
  \Gamma_{D2} = \Gamma_{012} \, .
\end{equation}
In the previous section (Equation~\eqref{eq:trap-Killing-spinors}) we
have found that the Killing spinors in the bulk are
\begin{equation}
  \begin{cases}
    \epsilon_L = \eu^{-\Phi/8} \left( \Id + \Gamma_{11} \right)
    \proj{flux}_- \exp [ \tfrac{1}{2} \phi_1 \Gamma_{45} +
    \tfrac{1}{2} \phi_2 \Gamma_{67} ] \epsilon_0 \, , \\
    \epsilon_R = \eu^{-\Phi/8} \left( \Id - \Gamma_{11} \right)
    \Gamma_u \proj{flux}_- \exp [ \tfrac{1}{2} \phi_1 \Gamma_{45} +
    \tfrac{1}{2} \phi_2 \Gamma_{67} ] \epsilon_1 \, .
  \end{cases}
\end{equation}
Using the commutation relations of $\proj{flux}_-$ with the components
of $\Gamma_u$, we find that the conservation of supersymmetry
requires
\begin{equation}
  \proj{flux}_- \left( \epsilon_0 - \frac{1}{\Delta} \Gamma_{1208} \, \epsilon_1 \right)
  + \proj{flux}_+ \left( \frac{m}{\Delta} \Gamma_{120} \left( \rho_1 \Gamma_5 - \rho_2 \Gamma_7
    \right) \right) \epsilon_1 = 0 \, .
\end{equation}
The two parts must vanish separately since the two projectors are
orthogonal\footnote{The contribution of the directions $x_1$ and $x_2$
  can be factored out, which is consistent with the fact that our
  configuration is T--dual to a D0--brane.}. We obtain the conditions
\begin{align}
  \rho_1 &= \rho_2 = 0 & \text{and} &&
  \epsilon_0 &= \Gamma_{1208} \, \epsilon_1 \, .
\end{align}
The first condition fixes the transverse position of the D2--brane, the
second one breaks half of the 16~supersymmetries, resulting in a total of
\textbf{8~preserved supercharges}:
\begin{equation}
  \boxed{  \begin{cases}
    \epsilon_L = \eu^{-\Phi/8} \, \Gamma_{1208} \left( \Id + \Gamma_{11} \right)
    \proj{flux}_-  \exp [ \tfrac{1}{2} \left( \phi_1 +
      \phi_2 \right) \Gamma_{67} ]  \epsilon_1 \, , \\
    \epsilon_R = \eu^{-\Phi/8} \left( \Id - \Gamma_{11} \right) \Gamma_u 
    \proj{flux}_-  \exp [ \tfrac{1}{2} \left( \phi_1 +
      \phi_2 \right) \Gamma_{67} ] \epsilon_1 \, .
  \end{cases}}
\end{equation}
It is convenient to introduce the Majorana--Weyl spinor $\wt \epsilon$,
\begin{equation}
  \label{eq:static-D2-spinors}
  \wt \epsilon = \eu^{-\Phi/8} \left( 1 + \Gamma_{11} \right) \proj{flux}_- \exp [ \tfrac{1}{2} \left( \phi_1 +
    \phi_2 \right) \Gamma_{67} ] \epsilon_1 \, ,
\end{equation}
and write the spinors conserved by the static embedding of the D2--brane as
\begin{align}
  \epsilon_L &= \Gamma_{1208} \, \wt \epsilon \, , & \epsilon_R = \Gamma_u
  \wt \epsilon \, .   
\end{align}
The spinor $\wt \epsilon$ is normalized such that it can written as the
sum of 8~orthogonal components\footnote{This is possible because $(\wt
  \epsilon)^T \, \wt \epsilon$ does not depend on $\phi_1$ or
  $\phi_2$.}:
\begin{equation}
  \wt \epsilon = \sum_{A = 1}^8 \wt \epsilon^A \, , \hspace{2em} ( \wt
  \epsilon^A )^T \, \wt \epsilon^B = \delta^{AB}  \, , \hspace{2em} A, B = 1, \dots, 8 \, ,
\end{equation}
where $\epsilon^T$ is the transposed spinor.

\subsection{BPS bound for the DBI action}
\label{sec:bps-excitations}

We have seen that the static embedding of the D2--brane into the fluxtrap breaks half of
the 16 supersymmetries of the bulk. Now we would like to describe a
different \textsc{bps} embedding that preserves only $1/4$ of the bulk
supersymmetries.

\paragraph{Hamiltonian formalism.}

In order better to understand the \textsc{bps} condition we pass to the
Hamiltonian formalism. The conjugate momentum to the bosonic variable
$X^\rho$ is given by
\begin{equation}
  P_\rho \equiv \frac{\del \mathcal{L}}{\del \dot X^\rho} = \mu_2
  \eu^{-\Phi} \frac{\dot X^\sigma g_{\rho\sigma}}{\sqrt{-\det g_{\alpha\beta}}} \, ,
\end{equation}
where $\dot X $ is the derivative with respect to $\zeta^0$:
\begin{equation}
  \dot X^\rho \equiv \frac{\del X^\rho}{\del \zeta^0} \, .  
\end{equation}
The Hamiltonian is therefore given by
\begin{equation}
  \label{eq:D2--bosonic-H}
  \mathcal{H} = P_\rho \dot X^\rho - \mathcal{L} = \mu_2 
  \eu^{-\Phi} \frac{\dot X^\rho \dot X^\sigma g_{\rho \sigma}}{\sqrt{-\det
      g_{\alpha\beta}}} + \mu_2  \eu^{-\Phi} \sqrt{-\det g_{\alpha\beta}} =
  \mu_2\frac{\eu^{-\Phi}}{\sqrt{-\det g_{\alpha\beta}}} \, ,
\end{equation}
and in particular the energy of the static embedding configuration
($\widehat g_{00} = -1$, $\rho_1 = \rho_2 = 0$) is
\begin{equation}
  \mathcal{H}^{\text{static}} = \frac{\mu_2}{g_3^2 \sqrt{\alpha'}}= \frac{1}{4 \pi^2 g_3^2 (\alpha')^2} \, .
\end{equation}

The last quantity we want to derive from the bosonic action is the \emph{angular
momentum} for a rotation in the direction of the Killing vector $V^\rho
\del_\rho = \del_{\phi_1}-\del_{\phi_2}$:
\begin{equation}
  \label{eq:D2--bosonic-J}
  \mathcal{J} = V^\rho P_\rho = \mu_2  \eu^{-\Phi}  \frac{V^\rho \dot
    X^\sigma g_{\rho\sigma}}{\sqrt{-\det g_{\alpha\beta}}} =
  \mu_2 \frac{\eu^{-\Phi}}{\sqrt{-\det g_{\alpha\beta}}} \frac{ \rho_1^2
    \dot \phi_1 - \rho_2^2 \dot \phi_2 }{1+ m^2 \left( \rho_1^2 + \rho_2^2 \right)} \, .
\end{equation}
Clearly, the angular momentum is vanishing for the static
configuration:
\begin{equation}
  \mathcal{J}^{\text{static}} = 0 \, .  
\end{equation}

\bigskip

In order to canonically quantize the \emph{fermionic} part of the action we
introduce the conjugate momentum
\begin{equation}
  \Pi_a \equiv \frac{\delta \mathcal{L}}{\delta \dot \psi^a} = \imath
  \frac{\mu_2 \eu^{-3\Phi/4}}{2 \sqrt{- \det g_{\alpha \beta}}} \psi^b
  \left( \Gamma_0 \hat \Gamma_0 \right)_{ba} \, , \hspace{2em} a,b = 1,
  \dots, 16 \, ,
\end{equation}
which by definition satisfies the canonical anticommutation relation
\begin{equation}
  \{ \Pi_a , \psi^b \} \equiv \imath \delta\du{a}{b} \, , \hspace{2em} a,b = 1, \dots, 16  
\end{equation}
whence
\begin{equation}
  \{ \Pi_a , \Pi_b \} = - \frac{\mu_2}{2} \frac{\eu^{-3\Phi/4}}{\sqrt{-
      \det g_{\alpha \beta}}} \left( \Gamma_0 \hat \Gamma_0 \right)_{ab} \, .
\end{equation}
Using the conjugate momentum one can directly write down the supercharges
that generate the supersymmetry transformations in
Equation~\eqref{eq:D2-fermionic-transformations}:
\begin{equation}
  Q_\epsilon = \imath \Pi_a \, \delta_\epsilon \psi^a  + \mathcal{O} \big( \text{(fermions)}^3 \big) \, ,
\end{equation}
which satisfy the anticommutation relation
\begin{equation}
  \{ Q^A, Q^B \} = \delta_{\epsilon^A} \psi^a \, \{ \Pi_a,
  \Pi_b \} \, \delta_{\epsilon^B} \psi^b 
  + \mathcal{O} \big( \text{(fermions)}^2 \big) \, .
\end{equation}
At this point we have all the ingredients to calculate the explicit
expression for the anticommutator, up to fermion bilinear terms. Since we want to consider
compare with the energy of the static embedding, we plug in the expressions for
the preserved Killing spinors in
Equation~\eqref{eq:static-D2-spinors}:
\begin{multline}
  \{ Q^A, Q^B \} = - \frac{\mu_2 \eu^{-3\Phi/4}}{2 \sqrt{ -\det
      g_{\alpha\beta}}} \left( \epsilon_L^A - \Gamma_{D2} \, \epsilon^A_R
  \right)^a \left( \Gamma_0 \hat\Gamma_0 \right)_{ab} \left( \epsilon_L^B - \Gamma_{D2} \,
    \epsilon_R^B \right)^b  \\
  = - \frac{\mu_2 \eu^{-3\Phi/4}}{2 \sqrt{ -\det g_{\alpha\beta}}}
  \left[ \Gamma_{1208} \, \wt \epsilon^A -
    \frac{\hat\Gamma_{012} \Gamma_u }{\sqrt{-\det g_{\alpha\beta}}} \wt
    \epsilon^A \right]^a \left( \Gamma_0 \hat\Gamma_0 \right)_{ab}
  \left[ \Gamma_{1208} \, \wt
    \epsilon^B-\frac{\hat\Gamma_{012} \Gamma_u }{\sqrt{-\det
        g_{\alpha\beta}}} \wt \epsilon^B \right]^b \, ,
\end{multline}
where we have dropped terms on the right-hand side containing two or more fermions,
leaving only the purely bosonic terms.

After a straightforward calculation we obtain a simple expression for the anticommutator:
\begin{equation}
  \{ Q^A, Q^B \} =  \left( \frac{\mu_2 \eu^{-\Phi}}{\sqrt{-\det
        g_{\alpha\beta}}} - \frac{\mu_2}{g_3^2 \sqrt{\alpha'}} \right) \delta^{AB} -
  \frac{\mu_2 \eu^{-\Phi}}{2 \sqrt{-\det g_{\alpha\beta}}} \{ \Gamma_8 -
  \frac{1}{\Delta} \Gamma_u, \hat \Gamma_0 \} ( \wt \epsilon^A)^T\Gamma_{08}\,
  \wt \epsilon^B \, .
\end{equation}
Using the explicit expression for $\Gamma_u$ in
Equation~\eqref{eq:Gamma-u}, and the pullback $\hat \Gamma_0$ in
Equation~\eqref{eq:Gamma0-hat}, we find that:
\begin{equation}
   \{ \Gamma_8 - \frac{1}{\Delta} \Gamma_u, \hat \Gamma_0 \} = -
  \frac{2m}{\Delta^2} \left(\rho_1^2 \dot \phi_1 - \rho_2^2 \dot \phi_2
  \right) \, ,
\end{equation}
which allows us to compare the anticommutator with the
expressions for the Hamiltonian and angular momentum that we have
found from the bosonic action in
Equations~\eqref{eq:D2--bosonic-H}~and~\eqref{eq:D2--bosonic-J}. The
final result is:
\begin{equation}
  \label{eq:D2-supercharge-anticommutator}
  \boxed{  \{ Q^A, Q^B \} = \left( \mathcal{H} - \mathcal{H}^{\text{static}}
  \right) \delta^{AB} + m \mathcal{J} \, ( \wt \epsilon^A)^T\Gamma_{08}
  \, \wt \epsilon^B }\, .
\end{equation}
The anticommutator vanishes for the static embedding since we are
discussing the supercharges preserved by this configuration. This is also the case for the \textsc{bps}--states that we construct in the following.

\paragraph{Rotating branes.}

The expression of the angular momentum suggests the following ansatz
for a rotating D2--brane:
\begin{align}
  \phi_1 &= \dphi \, \zeta^0 \, , & \phi_2 &= - \dphi \, \zeta^0 \, ,
\end{align}
where $\dphi$ is constant and all the other transverse coordinates
have a fixed value independent of $\zeta^0$.  The non-trivial
pullbacks of metric and gamma matrices are given by
\begin{align}
  \widehat g_{00} &= - \frac{ 1 + \left( \rho_1^2 + \rho_2^2 \right) \left(
       m^2 - \dphi^2 \right) } {1+ m^2 \left( \rho_1^2 + \rho_2^2
    \right)} \, ,  \\
  \hat \Gamma_0 &= \Gamma_0 - \frac{\dphi}{m} \Gamma_8 +
  \frac{\dphi}{m \Delta} \Gamma_u \, .
\end{align}
The bosonic part of the Lagrangian is then given by
\begin{equation}
  \mathcal{L}_b = -\frac{1}{4 \pi^2 g_3^2 (\alpha')^2} \sqrt{ 1 +
    \left( \rho_1^2 + \rho_2^2 \right)\left( m^2 - \dphi^2 \right)} \,  .
\end{equation}
\textsc{bps} states are extrema of the action. The non-trivial
BPS equations are:
\begin{equation}
  \rho_1 \left( m^2 - \dphi^2 \right) = \rho_2 \left( m^2 - \dphi^2 \right) = 0 \,.
\end{equation}
These are satisfied either if $\rho_1 = \rho_2 = 0 $, which is the
static embedding, or if
\begin{equation}
  \label{eq:rotating-D2-trap}
  \dphi = \pm m \, .
\end{equation}
This is the \emph{rotating D2--brane embedding}.  Note that we have \emph{not}
restricted to small fluctuations about the static brane: Even if we
are not in a linear approximation, the frequency is independent of the amplitude and no
conditions are imposed on the position of the D2--brane in $\rho\ll{1,2}$ or the other
transverse directions.

By substituting the condition in Equation~\eqref{eq:rotating-D2-trap}
into the general expressions for energy and angular momentum we
find
\begin{align}
  \mathcal{H}^{\text{rot}}_\pm &= \frac{1}{4 \pi^2 g_3^2 (\alpha')^2}
  \left( 1 + m^2 \left( \rho_1^2 + \rho_2^2 \right) \right) &
  \mathcal{J}^{\text{rot}}_\pm = \pm  \frac{m}{4 \pi^2 g_3^2
    (\alpha')^2} \left( \rho_1^2 + \rho_2^2 \right) \, ,
\end{align}
where the $\pm$ refers to the two branches of the solution $\dphi
= \pm m$.  After subtracting the energy of the static configuration we
obtain the \textsc{bps} condition
\begin{equation}
  \frac{\mathcal{H}^{\text{rot}}_\pm -
    \mathcal{H}^{\text{static}}}{\mathcal{J}^{\text{rot}}_\pm} = \pm m \, .
\end{equation}
In order to verify that this is indeed a bound, we have to turn to the
fermionic action. A Killing spinor $\epsilon$ is conserved iff:
\begin{equation}
  \epsilon_L = \Gamma_{D2} \, \epsilon_R \, . 
\end{equation}
Using the expression for the bulk Killing spinors in
Equation~\eqref{eq:trap-Killing-spinors}, and the pullback of the
$\Gamma$ matrices on the rotating brane ansatz, we obtain the
equation
\begin{equation}
  \proj{flux}_- \left( \epsilon_0 - \Gamma_{1208} \, \epsilon_1 \right)
  + \proj{flux}_+ \left( m \, \Gamma_{120} \left( \Id \pm
      \Gamma_{08} \right) \left( \rho_1 \Gamma_5 - \rho_2 \Gamma_7
    \right) \right) \epsilon_1 = 0 \, .
\end{equation}
The two parts must vanish separately since the two projectors are
orthogonal. This implies
\begin{align}
  \epsilon_0 &= \Gamma_{1208} \, \epsilon_1 \, , & \text{and} &&
  \left( \Id \pm \Gamma_{08} \right) \epsilon_1 &= 0 \, .
\end{align}
The two conditions together preserve a total of
\textbf{4~supercharges}. Explicitly:
\begin{equation}
  \boxed{\begin{cases}
    \epsilon_L = \eu^{-\Phi/8} \, \Gamma_{1208} \left( \Id + \Gamma_{11} \right)
    \proj{flux}_-  \exp [ \tfrac{1}{2} \left( \phi_1 +
      \phi_2 \right) \Gamma_{67} ] \left( \Id \mp \Gamma_{08} \right) \epsilon_2 \, , \\
    \epsilon_R = \eu^{-\Phi/8} \left( \Id - \Gamma_{11} \right) \Gamma_u 
    \proj{flux}_-  \exp [ \tfrac{1}{2} \left( \phi_1 +
      \phi_2 \right) \Gamma_{67} ] \left( \Id \mp \Gamma_{08} \right) \epsilon_2 \, ,
  \end{cases}}
\end{equation}
where $\epsilon_2$ is a constant Majorana spinor.  This is precisely
the same form that we have found for the static embedding in
Equation~\eqref{eq:static-D2-spinors}, with an extra projector $\frac{1}{2}\left(
  \Id \mp \Gamma_{08} \right) $.  To be precise, a given Killing
spinor $\epsilon = \left( \Gamma_{1208} + \Gamma_u \right) \wt
\epsilon $ which is preserved in the static embedding is also
preserved by the rotating brane if
\begin{equation}
    \Gamma_{08} \, \wt \epsilon_{\text{pres}} = \mp \wt \epsilon_{\text{pres}} \, , 
\end{equation}
and not preserved otherwise: $ \Gamma_{08} \, \wt
\epsilon_{\cancel{\text{pres}}} = \pm \wt
\epsilon_{\cancel{\text{pres}}} $.

Using the expression for the anticommutator of the supercharges in
Equation~\eqref{eq:D2-supercharge-anticommutator} we find that the
supercharges corresponding to Killing spinors preserved and not
preserved by the rotating D2--brane satisfy
\begin{align}
  \{ Q^A_{\text{pres}}, Q^B_{\text{pres}} \} &= \left(
    \mathcal{H} - \mathcal{H}^{\text{static}} 
  \mp m \mathcal{J} \right) \delta^{AB}\\
  \{ Q^A_{\text{pres}}, Q^B_{\cancel{\text{pres}}} \} &= 0 \\
  \{ Q^A_{\cancel{\text{pres}}}, Q^B_{\cancel{\text{pres}}} \} &=
  \left( \mathcal{H} - \mathcal{H}^{\text{static}} \pm m \mathcal{J}
  \right) \delta^{AB} \, .
\end{align}
This implies that
\begin{align}
  \left( \mathcal{H} - \mathcal{H}^{\text{static}} \right) + m \mathcal{J} &\ge 0 \,
  , & \left( \mathcal{H} - \mathcal{H}^{ \text{static}}  \right) - m \mathcal{J} &\ge 0
  \, ,
\end{align}
where one of the two conditions is trivial depending on the sign of $m
\mathcal{J}$. The bound is saturated if $ \dphi = \pm m$:
\begin{equation}
  \mathcal{H}^{\text{rot}}_\pm - \mathcal{H}^{\text{static}} 
  \mp m \mathcal{J}^{\text{rot}}_\pm = 0 \, . 
\end{equation}

\subsection{Low energy effective gauge theory}
\label{sec:low-energy-effective}


In this section we derive the low energy action describing the
dynamics of the D2--brane in the fluxtrap background. The parameter
$m$ that we have introduced in the identifications in
Section~\ref{sec:fluxtrap} will appear explicitly as a real mass term for
the fields describing the motion of the D2--brane in the directions
$x_4 \dots x_7 $.

Let us start with the kappa--symmetry-fixed \textsc{dbi} action at
second order in the fermions. In order to get the canonical
normalization for the fermionic term it is convenient to pass to the
democratic formulation (see~\cite{Martucci:2005rb}) in which the
action is written as:
\begin{equation}
  S = - \mu_2 \int \di^3 \zeta \; \eu^{- \Phi} \sqrt{ - \det (g_{\alpha
        \beta} + B_{\alpha \beta} ) }  \left[ 1 - \frac{1}{2} \bar \psi \left(
      (g + B)^{\alpha \beta} \Gamma_\beta D_\alpha + \Delta^{(1)} \right) \psi
  \right] \, ,
\end{equation}
where
\begin{align}
  D_\alpha &= \del_\alpha X^\mu \left(\nabla_\mu + \frac{1}{8} H_{\mu
      m n} \Gamma^{m n} \right) \, ,\\
  \Delta^{(1)} &= \frac{1}{2} \Gamma^m \del_m \Phi - \frac{1}{24}
  H_{mnp} \Gamma^{mnp} \, .
\end{align}
Since we are only interested in the low energy dynamics we expand all
the terms at the respective leading order. The bulk fields are
\begin{align}
  g_{\mu \nu} \di X^\mu \di X^\nu &= \di \vec x^2_{0 \dots 9} +
  \mathcal{O}(X^4) \, , \\
  H_{\mu \nu \rho} \di X^\mu \wedge \di X^\nu \wedge \di X^\rho &= 2 m
  \left( \rho_1 \di \rho_1 \wedge \di \phi_1 - \rho_2 \di \rho_2
    \wedge \di \phi_2 \right) \wedge \di x_8 + \mathcal{O} ( X^5 ) \, , \\
  \eu^{-\Phi} &= \frac{1}{g_3^2 \sqrt{\alpha'}} \left( 1 + \frac{m^2}{2}
    \left( \rho_1^2 + \rho_2^2 \right) \right) + \mathcal{O} (X^4) \, .
\end{align}
In our consistent truncation the dynamics only depends on $\zeta^0$, hence
\begin{equation}
  \sqrt{- \det ( g_{\alpha \beta} + B_{\alpha \beta} ) } = 1 -
  \frac{1}{2} \dot X^\sigma \dot X_\sigma + \mathcal{O} (X^4) \, ;
\end{equation}
moreover the only relevant covariant derivative, at leading order
reduces to
\begin{equation}
  \nabla_0 = \del_0 \, .
\end{equation}
A straightforward calculation shows that
\begin{align}
  g^{\alpha \beta} \Gamma_\beta D_\alpha &= - \Gamma_0 \del_0 +
  \mathcal{O} (X^2) \, ,\\
  \Delta^{(1)} &= - \frac{m}{2} \left( \Gamma_{45} - \Gamma_{67} \right)
  \Gamma_8 + \mathcal{O} ( X^2) \, .
\end{align}
And substituting the expansion into the action we obtain:
\begin{equation}
  S = %
  - \frac{1}{8 \pi^2 g_3^2 (\alpha')^2}\int \di^3 \zeta \left[ -
    \dot X^\sigma \dot X_\sigma + m^2 \left( \rho_1^2 + \rho_2^2
    \right) + \bar \psi\,
  \Gamma_0 \dot \psi + \frac{m}{2} \bar \psi \left( \Gamma_{45} -
    \Gamma_{67} \right) \Gamma_8 \psi \right] + \dots \, .
\end{equation}
The result is more transparent in rectilinear coordinates,
\begin{align}
  z_1 &= x_4 + \imath x_5 = \rho_1 \eu^{\imath \phi_1} \, ,& z_2 &= x_6 +
  \imath x_7 = \rho_2 \eu^{\imath \phi_2} \, ,
\end{align}
in which the relevant part of the action becomes
\begin{equation}
  S = \frac{1}{8 \pi^2 g_3^2 (\alpha')^2}\int \di^3 \zeta \left[ 
    \dot z_1 \dot {\bar z}_1 + \dot z_2 \dot {\bar z}_2 - m^2 \left( z_1
      \bar z_1 + z_2 \bar z_2 \right) - \bar \psi\, \Gamma_0 \dot \psi -
    \imath m \, \bar \psi \left( \proj{$z_1$}_- -
      \proj{$z_2$}_- \right) \Gamma_8 \psi \right]  ,
\end{equation}
where the projectors \proj{$z_1$} and \proj{$z_2$} are defined by
\begin{align}
  \proj{$z_1$}_- &= \frac{1}{2} \left( \Id - \imath \Gamma_{45}
  \right) \, ,& \text{and} &&
  \proj{$z_2$}_- &= \frac{1}{2} \left( \Id - \imath \Gamma_{67} \right)
  \, .
\end{align}

With gauge fields included, the quadratic action contains the additional term
\begin{equation}
  S\ll{\text{gauge}} = - \frac{1}{4 g_3^2} \int \di \uu 3 \zeta \,
  F\ll{\alpha \beta} F\uu{\alpha \beta} \, .
\end{equation}

For $N$ identical D2--branes in the fluxtrap geometry, the $U(1)$ gauge connection is
promoted to a $U(N)$ gauge connection, the scalar fields and fermions are
promoted to matrices in the adjoint representation of $U(N)$, and the action
is replaced by a single-trace version of itself, which is uniquely determined up
to commutator terms.  Since commutator terms involve at least three fields
and we are only working to quadratic order, such terms do not affect the properties
of the twisted mass in the gauge theory.

\section{Relation to the gauge/Bethe correspondence}
\label{sec:gaugebethe}

\subsection{Brane configuration without twisted masses}

We will now make some brief comments on the connection of
our work to the gauge/Bethe correspondence of~\cite{Nekrasov:2009uh,Nekrasov:2009ui}.
The string theory embedding of the gauge theories there involve
D2--branes suspended between NS--fivebranes, with D4--branes added to the background, as shown in
Table~\ref{tab:NS5-embedding}.
This brane configuration is a subset of the one described
in~\cite{Orlando:2010uu}, with the NS5'--brane removed.  
After the removal of the orthogonal NS5' and the
exchange $x\uu 2\leftrightarrow x\uu 6$ and $x\uu {3,4,5} \leftrightarrow x\uu {7,8,9}$,
the D2, D4 and NS5 branes of \cite{Orlando:2010uu} becomes the branes we consider here.  The D4--branes here are located at an arbitrary position
between the two NS fivebranes.

Prior to the addition of the twisted mass deformation, the
configuration in Table~\ref{tab:NS5-embedding} preserves $(4,4)$
supersymmetry in 1+1 dimensions~\cite{Brodie:1997wn}, unlike the configuration of~\cite{Orlando:2010uu} which preserves only (2,2), due to the presence of the orthogonal NS5'--brane that was used to give an infinite
mass in the superpotential for the adjoint chiral multiplet degrees of
freedom, following~\cite{Hanany:1997vm,Witten:1997ep}.  In the configuration of~\cite{Orlando:2010uu}, the presence of the NS5'--brane as one of the
two boundaries for the D2 leaves only two massless adjoint scalars,
enough to fill out a (2,2) vector multiplet.  In our current
configuration, by contrast, prior to the fluxtrap deformation, there
are four massless scalars in the adjoint, enough to fill out a vector
multiplet of (4,4) supersymmetry.  The $A\ll 2$ component of the
D2--brane gauge field obeys Dirichlet boundary conditions at the
NS--fivebrane~\cite{Hanany:1996ie}, so there are no additional massless bosonic
2-dimensional degrees of freedom in the adjoint, beyond those of the
$(4,4)$ vector multiplet, consisting of the 2D gauge connection, and motions in
the $x\ll{6,7,8,9}$ directions.

The two NS--fivebranes are separated in the $x\ll 2$ direction by a distance $\d\ll 2$, and
there is a set of $L$ D4--branes touching the N D2s.  For any such configuration, there is
a set of massless hypermultiplets in the fundamental and in the antifundamental
representation of $SU(N) $, consisting of open strings connecting the D2--branes and the
D4--branes~\cite{Brodie:1997wn,Hanany:1997vm}.  The $(4,4)$ supersymmetry forces an interaction~\eqref{superpot44} 
which in ${\cal N} = (2,2)$ language is a cubic superpotential involving the fundamental
chiral multiplets, the antifundamental chiral multiplets, and the adjoint chiral multiplet:
\begin{equation}
  W = \wt{{\cal Q}} \cc \phi \cc {\cal Q} \ . 
\end{equation}

\subsection{Fluxtrap deformation of the brane configuration}

Thus the configuration we have described reproduces exactly the gauge theory of \cite{Nekrasov:2009uh,Nekrasov:2009ui}, with precisely one exception: The sole missing ingredient is the twisted mass deformation for
chiral multiplets in the fundamental and adjoint representations.
 These mass terms
are present in~\cite{Nekrasov:2009uh,Nekrasov:2009ui}, playing a key role in the infrared dynamics.  So we now deform our brane configuration by  
the fluxtrap deformation of the closed string
background, which 
adds a
twisted mass deformation for the adjoint and chiral multiplets of the D2--brane gauge theory ,  breaking SUSY to ${\cal N} = (2,2)$.  

The twisted mass for the adjoint in 1+1 dimensions simply descends from a local term in 2+1 dimensions, the real mass for the adjoint chiral multiplet.  To see that the twisted mass
term for the fundamental and antifundamental chiral multiplets must be present, one need
only verify that the deformation preserves ${\cal N} = (2,2)$ supersymmetry, and note
that the superpotential must be neutral under the symmetry operator $Z$ defining the
central charge.  Since the adjoint chiral multiplet is not neutral, the fundamental and
antifundamental chiral multiplets must be non-neutral as well.  Together they must cancel
the $Z$-charge of the adjoint chiral multiplet, and so each must have a Z-charge equal
to $-\hh$ the Z-charge of the adjoint chiral multiplet.  Thus the fundamental matter is
forced to have a mass equal to exactly half that of the adjoint matter~\cite{Nekrasov:2009uh}.

It remains to demonstrate that the fluxtrap deformation can be combined consistently
with the presence of these other branes,
the NS5s and D4s.  Once this is shown, our string realization of the gauge theories of~\cite{Nekrasov:2009uh,Nekrasov:2009ui} will be complete.

One might have questioned whether the ingredient added here -- the twisted mass
deformation for the chiral multiplets \emph{via} the fluxtrap solution -- can in fact be combined consistently with the other
ingredients of~\cite{Orlando:2010uu}, the NS5--branes on which the D2's
terminate, and the D4--branes providing the matter in the fundamental
representation.  The answer to that question is affirmative.  The D4--branes can be
added unproblematically to the solutions described in this article;
the string coupling $\exp[\Phi]$ is bounded
above by an arbitrarily small value in all the solutions we consider, the
backreaction of the D4--branes on the rest of the geometry can
be made arbitrarily small.

At first sight, combining the NS5's with the fluxtrap deformation may
appear to be a trickier  issue.  Both the NS5's and the fluxtrap are solutions of nonlinear equations of motion of the 
massless modes of closed string theory.  There is no
principle that guarantees that such solutions need superpose with one another.  However
we show by explicit construction, in Appendix~\ref{sec:ns5-fluxtrap}, that the solutions do in fact combine; 
there is a combined fivebrane-fluxtrap solution that reduces to the pure fluxtrap when the
fivebranes are moved to infinity, and reduces to a solution of arbitrarily positioned parallel
NS--fivebranes when the fluxtrap deformation is turned off.

 In Appendix~\ref{sec:ns5-fluxtrap} we have written the full solution
 for the fluxtrap deformation of the geometry
of parallel (but not necessarily coincident) NS--fivebranes, which break the supersymmetry 
of the fluxbranes again by half.
This demonstrates the consistency of combining the NS--fivebranes with the
fluxbrane deformation, establishing our construction as an exact string solution
reducing to the Nekrasov--Shatashvili gauge/Bethe system at low energies.

 In the Appendix we have for completeness also included
 solutions of the Dirac--Born--Infeld action of the D2--branes in the 
 fivebrane-fluxbrane background, which shows the persistence of the
 exact BPS formula for rotating trajectories
 of the adjoint fields with twisted masses, in the presence
 of the NS--fivebranes.  We construct static and rotating \textsc{bps} solutions
of the D2--brane DBI action in the NS5--fluxbrane background, which preserve 
4 and 2 supercharges, respectively, and again satisfy the relation $E- E_0 = 
\abs{m\cc J}$.  
These static and rotating solutions are classically \textsc{bps}--saturated embeddings of D2--branes into the fivebrane-fluxtrap geometry. The embeddings exactly saturating the \textsc{bps} bound
 $E - E\ll 0 = \abs{\cc m J \cc }$, can be thought
of as representing a condensate of \textsc{bps}-saturated particles in $(2+1)$--dimensional gauge
theory on the interval.

\subsection{Quantum nonabelian symmetry from the brane construction}
\label{sec:quantum-nonabelian}

The existence of this brane construction has the potential to teach us many interesting things
about the remarkable relationships among two-dimensional gauge theories.  To take
one immediate example, we will examine the emergence of nonabelian global
symmetries relating the Nekrasov--Shatashvili gauge theories with different ranks $N
= \#\cc {\rm D2}$.
Take the case of two NS--fivebranes, parallel
and separated by a distance $\d\ll 2$, with $N$
D2--branes suspended between them and a twisted mass parameter $m$ characterizing
the strength of the fluxtrap, which gives the adjoint chiral multiplets a twisted mass
$m$ and the fundamental and antifundamental chiral multiplets a twisted mass
${m\over 2}$.

As the separation $\d\ll 2$ between the fivebranes is taken to zero with $g\ll 3$ held
fixed, the two-dimensional gauge coupling becomes infinitely strong and quantum
effects dominate the system; there is a rich set of quantum vacuum states depending
on $N$ and $L$, which have been
shown~\cite{Nekrasov:2009uh,Nekrasov:2009ui}  to be in one-to-one
correspondence  with the \emph{full} Hilbert space of the $N$--magnon
sector of a spin chain with $L$ spin sites.
The set of vacuum states unexpectedly arranges itself to respect a global
$SU(2)$ symmetry~\cite{Orlando:2010aj} organizing the states into $SU(2)$ representations with irreducible
components of dimension at most $L+1$.  As $\d\ll 2\to 0$ this symmetry becomes an
exact symmetry of the supersymmetric vacuum states of the gauge theories.

From the spin chain point of view, the $SU(2)$ is immediately apparent: each spin
variable transforms in a two-dimensional representation and the full state of the system
is trivially a tensor product of those.  From the gauge theory point of view, on the
other hand, the quantum $SU(2)$ has been mysterious.  Particularly 
striking is the nature of the action of the SU(2) generators on the quantum number $N$,
the number of D2--branes.  In the correspondence of~\cite{Nekrasov:2009uh,Nekrasov:2009ui}, the
number $N$ is ${L\over 2}$ plus the Cartan generator of the $SU(2)$.  The raising
and lowering operators of the $SU(2)$ therefore raise and lower the rank
of the $(1+1)$--dimensional gauge theory itself.  This peculiar SU(2) is clearly
a powerful and unfamiliar type of symmetry -- it acts not on the \textsc{susy} vacuum
sector of a particular gauge theory, but on the set of \textsc{susy} vacua of an \emph{ensemble}
of gauge theories of different ranks $N$, mapping vacua of theories 
of different rank to one another.  

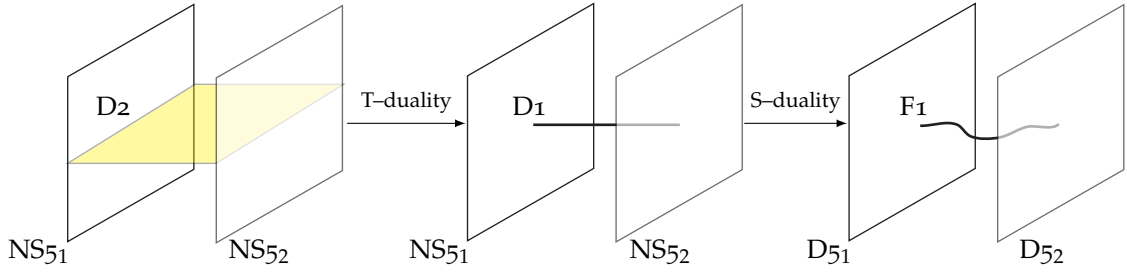
\begin{figure}
  \centering
  \begin{tikzpicture}
    \matrix (m) [matrix of nodes, column sep=.7cm]
    { 
      \draw (1,2.1) node[anchor=north east]{\small D2};  
      \draw (0.2,0.2) node[anchor=north east]{\small NS5$_1$};
      \draw (3.1,0.2) node[anchor=north east]{\small NS5$_2$};
      \node [inner sep=0pt,above right] {\includegraphics[scale=.5]{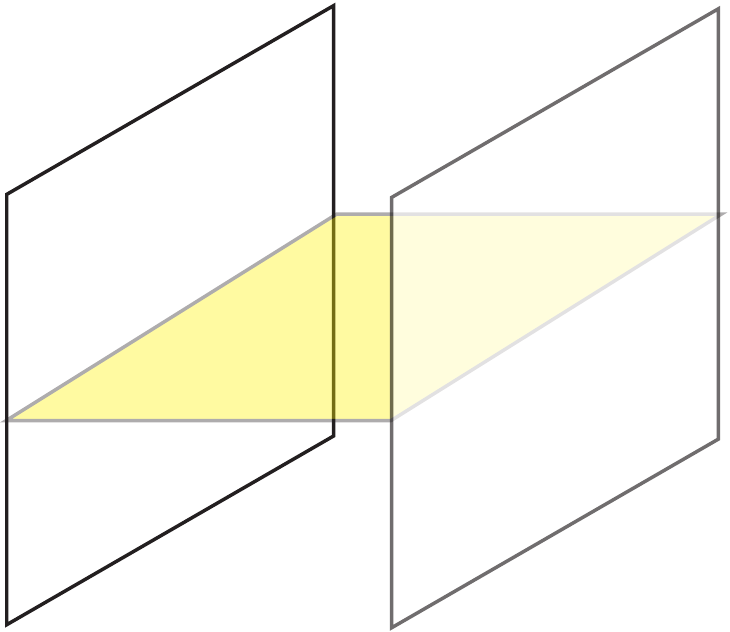}};
      & 
      \draw (1.2,2.1) node[anchor=north east]{\small D1};  
      \draw (0.2,0.2) node[anchor=north east]{\small NS5$_1$};
      \draw (3.1,0.2) node[anchor=north east]{\small NS5$_2$};
      \node [inner sep=0pt,above right] {\includegraphics[scale=.5]{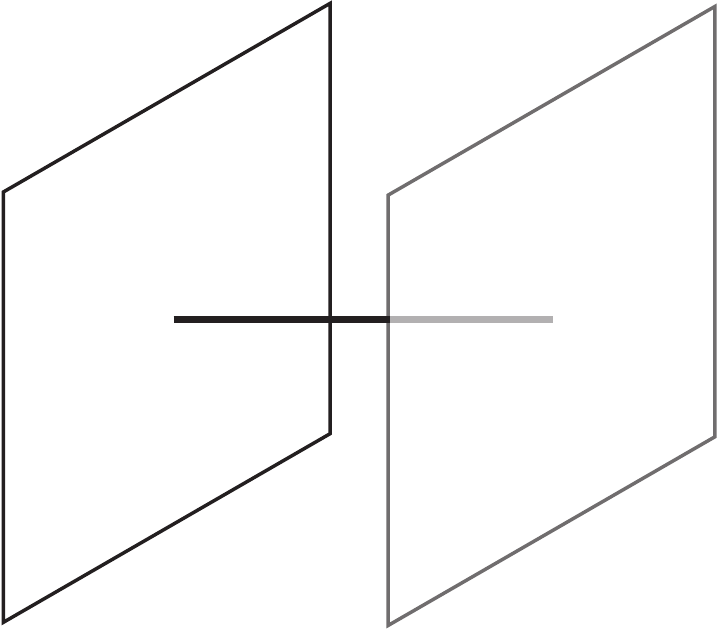}};
      & 
      \draw (1.2,2.1) node[anchor=north east]{\small F1};  
      \draw (0.2,0.2) node[anchor=north east]{\small D5$_1$};
      \draw (3,0.2) node[anchor=north east]{\small D5$_2$};
      \node [inner sep=0pt,above right] {\includegraphics[scale=.5]{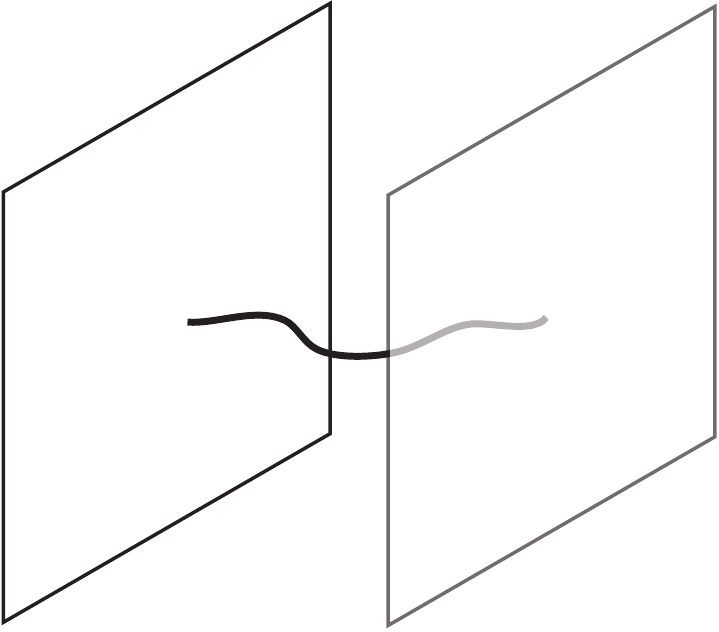}};
      \\};
    \path[>=latex,->,font=\scriptsize] (m-1-1) edge node[auto] {T--duality} (m-1-2);
    \path[>=latex,->,font=\scriptsize] (m-1-2) edge node[auto] {S--duality} (m-1-3);
 \end{tikzpicture}
  \caption{Realization of the $SU(2)$ symmetry via dualities in string
    theory.}
  \label{fig:U-duality}
\end{figure}

The string embedding sheds some light on the origin of this mysterious
$SU(2)$ symmetry.  Upon compactification of the spatial direction of the 
$(1+1)$--dimensional gauge theory, the brane configuration becomes equivalent
under T--duality to a system of D1--branes suspended between NS--fivebranes
in type IIB string theory.  It is well known that this configuration supports
a dynamical $SU(2)$ gauge symmetry propagating on the system of NS--fivebranes
that is broken when the NS5s are separated and restored when the NS5s become
coincident.
This gauge symmetry has the property that the number of D1--branes suspended
between NS5s does indeed play the role of the Cartan generator, with the
raising and lowering operators literally creating and destroying D1--branes.
This seemingly exotic action of the gauge symmetry can be understood most
simply through the S-duality of type IIB string theory, under which the NS5--branes
become D5-branes and the D1--branes suspended between them become 
open fundamental strings, transforming in the adjoint of the $SU(2)$
(see Figure~\ref{fig:U-duality}).

%

Though a gauge symmetry from the point of view of the fivebranes themselves,
the $SU(2) $ appears to the D2--branes  as a global symmetry,  because the 
gauge bosons of the $SU(2)$ propagate on the fivebranes and not on the twobranes.
This $SU(2)$ also becomes unbroken when the NS5s become coincident.   It appears 
to pass all the most obvious tests to play the role of the $SU(2)$ symmetry organizing
the ground states of the supersymmetric gauge theories of~\cite{Nekrasov:2009uh,Nekrasov:2009ui}.
Indeed, the string-theoretic embedding opens an even more surprising possibility:
if the $SU(2)$ is an exact dynamical symmetry of the system, then it ought to act on non-vacuum
as well as vacuum states.  This suggests an even more remarkable set of relationships
among two-dimensional gauge theories and their quantum states than that
contemplated in~\cite{Nekrasov:2009uh,Nekrasov:2009ui}.

There are several facts that may somewhat temper one's hopes in this regard.  Among
them:
\bi
\item{It is not clear that there is a decoupling limit in which the dynamics of
non-vacuum states of the twobranes decouple from the degrees of freedom of 
the bulk and from the fivebranes.  The 
$SU(2) $ may be an exact symmetry of the system of string theory on the fivebranes, but
that symmetry may not act on states that can be understood as excitations of the
D2--branes alone; it may be necessary to add states in the bulk or attached to the
fivebranes, in order to fill out complete $SU(2)$ representations.}
\item{The $SU(2)$ may be destroyed at the quantum level by the fluxtrap deformation.
The fluxtrap deformation does not carry $SU(2)$ quantum numbers and 
cannot break the gauge symmetry explicitly; nor can it trigger spontaneous breaking
in the usual sense, as the solution with coincident fivebranes is still a valid
supersymmetric solution even after the fluxtrap deformation (as shown in
Appendix~\ref{sec:ns5-fluxtrap}).  Rather, the danger is that the fluxtrap deformation may trigger quantum
dynamics on the fivebranes that would give rise to confinement of the $SU(2)$.  The
undeformed six-dimensional theory certainly does \emph{not} confine, and indeed
confinement
 would be impossible in a fully Poincar\'e-invariant six-dimensional gauge theory.  However the
fluxtrap breaks some of the Poincar\'e symmetry of the six-dimensional theory: 
 in
order to preserve supersymmetry and allow the D2--branes to be suspended between
them while preserving supersymmetry,
the NS5s must be oriented in the $x\uu{0,1,6,7,8,9}$ directions.
The fluxtrap deformation therefore breaks the Poincaré symmetry 
on the NS--fivebranes down to $SO(2,1)\ll{019} \times SO(2) \ll{67}$, times
translational symmetries in the $0,1,2,$ and $8$ directions.  The supersymmetry of
the fivebrane theory is also partially broken by the fluxtrap down to eight supercharges, which is
a low enough amount to allow a gauge coupling running to strong at long distances.
This peculiar, Lorentz-breaking six-dimensional gauge theory is sufficiently unfamiliar that 
we cannot rule out the possibility that the deformed theory may confine in the infrared.  It is a logical possibility that strong coupling
dynamics deforms the moduli space such that there is no point at which
the $SU(2)$ is restored, perhaps similarly to~\cite{Seiberg:1994rs}.  Our attitude in the
present discussion is to take the non-confinement
of the $SU(2)$ at the quantum level as a working hypothesis, but by no means a proven
fact.}
\item{Even if our string embedding explains the emergence of an $SU(2)$ 
after compactification of $x\ll 1$, we still have not explained the apparent existence of
an $SU(2)$ prior to compactification, which still appears to be valid.  The uncompactified
type IIA NS--fivebrane does not support a gauge symmetry in the usual sense.}
\ei
The points above make clear that the string embedding offers a plausible framework for 
explaining the $SU(2)$ symmetry but not a full explanation in the absence of further 
refinement.

All comments above apply, on the gauge theory side, the spin-chain side, and 
the string theory side, to the case of $k $ fivebranes, where the gauge theory becomes a
more complicated quiver~\cite{Nekrasov:2009uh,Nekrasov:2009ui}, the brane construction
has $k$ parallel fivebranes with twobranes suspended between them,
 and the mysterious quantum symmetry of the gauge/spin-chain is enhanced to $SU(k)$,
 as reproduced by the dynamics of the $k$ fivebranes.

 \section{Conclusions}
 \label{sec:conclusions}
 
 In this paper we have constructed a simple solution of type II string theory, the fluxtrap
solution, realized
as a T--dual of a free quotient of flat space preserving half the supersymmetry of
the flat covering space.  This fluxtrap can be viewed as a lift of the $\O$--deformation
to string theory.  This background unifies the $\O$--deformed 4-dimensional gauge theories of
\cite{Nekrasov:2003rj} and the Lorentz-invariant 3-dimensional gauge theories with
twisted masses~\cite{Nekrasov:2009uh,Nekrasov:2009ui}; each gauge theory is realized on a type
of D--brane in the fluxtrap background, with the former oriented longitudinal to
the $z_{1,2}$ directions of the fluxtrap geometry, and the latter transverse to
$z_{1,2}$.  The coupling to the curved metric, $B$--field and dilaton gradient 
of the closed string background provide simple ways of understanding the deformed
dynamics of each type of theory.    In particular we saw explicitly that the same
deformation of the closed string background that produces the twisted masses on
a set of D2--branes transverse to the fluxtrap geometry, can also produce the
Omega deformation of the gauge theory on 
a set of Euclidean D3'--branes longitudinal the fluxtrap geometry.

We have constructed the \textsc{bps}-saturated classical solutions 
of the D2--branes rotating in the fluxtrap background.  These states are half-supersymmetric
states of the branes in the fluxtrap background, preserving 4 of the 8 dynamical
supercharges preserved by the static brane and satisfying the exact relation $E- E_0 = 
\abs{m\cc J}$ where $m$ is the twisted/real mass parameter and $J$ is the angular momentum
generator that rotates $z\ll 1$ and $z\ll 2$ with opposite phases.  The translationally
invariant classical solutions are simply Bose-Einstein condensates of \textsc{bps} oscillators
that have zero momentum in the $x_{1,2}$ direction.   We have further shown
in the appendix that these classical solutions have analogous \textsc{bps} solutions when 
NS--fivebranes are added to the background,  with the D2s suspended between the NS--fivebranes, and either rotating or not.

We have discussed (without much detail) the addition of D4--branes together with NS--fivebranes
to the solution, in order to make contact with the gauge/Bethe correspondence of
Nekrasov and Shatashvili.   By doing this, we have found a partial explanation
of the mysterious quantum mechanical $SU(k)$ symmetry that acts on 
the quantum ground states of the system when the two-dimensional gauge coupling goes
to infinity.  Certain gaps, however, remain in this explanation.

The emergence of (not necessarily normalizable) classical \textsc{bps} states, consisting of excitations of the 
$z\ll 1$ and $\bar{z}\ll 2$ degrees of freedom, and their superpartners,
is intriguing.  The \textsc{bps} formula for these states suggests that their energies do 
not become infinite even when the two-dimensional gauge coupling goes to
infinity.  It would appear that the quantum vacuum states of the gauge theories
are augmented by a set of non-vacuum \textsc{bps} states that survive and should organize themselves into SU(k) representations
(in the presence of $k$ fivebranes)  in the strong
coupling limit.  It would be interesting indeed to understand how the spin chain
picture could be enlarged to understand these states.

It may seem puzzling why such  a simple deformation as a
dimensional reduction on a twisted circle should need to be
understood in terms of a complicated-looking supergravity solution
involving curved metrics, $B$--fields and dilaton gradients.  And yet 
already we have seen that some of these dynamical elements have allowed us
to see aspects of the $\O$--deformation of 4D gauge theory, and twisted mass deformation
of 3D and 2D gauge theory, with a certain clarity.  Universal principles counsel that it is
always better to use a description where irrelevant heavy degrees of freedom have been
removed from the system.  The irrelevant degrees of freedom, which were the
momentum modes on the $x\uu{\wt{8}}$ circle, have been transformed in the
T--dual picture into infinitely heavy winding string modes, which play
no role in the dynamics.  Finally we would like to note that the fluxtrap background
represents an integrable string theory on general grounds, as it is
equivalent under a T-duality to a free quotient of flat space.  Any such background
is solvable by a generally applicable recipe \cite{Hellerman:2006tx}
and indeed this particular background has already been to some extent
solved, in its description as a fluxbrane, in \cite{Takayanagi:2001jj}.

 We consider it likely that the fluxtrap description of the
$\O$--background will prove efficient for computations where the description as
a twisted compactification is unwieldy.  There is hope that this solution will further
the investigation of the remarkable relationships among gauge theories first
noted in~\cite{Nekrasov:2009uh,Nekrasov:2009ui}.

The results of this article were announced in
a talk at the ``Branes and Bethe Ansatz in Supersymmetric Gauge Theory Workshop'', March, 2011~\cite{DomenicoTalk}.

\subsection*{Acknowledgements}

It is our pleasure to thank Luca Martucci, Konstadinos Sfetsos, and Linda Uruchurtu  for correspondence.

D.O. and S.R. are happy to thank the \emph{National Center for Theoretical
Sciences, Taipei, Taiwan R.O.C.} for hospitality during the final
stages of this work.  S.H. would like to thank the theory groups at 
\emph{CERN}, \emph{ETH Zurich}, \emph{LPT Ecole Normale Superieure}, and \emph{AEI Potsdam}
for hospitality while this work was in progress.

This work was supported by the World Premier International Research
Center Initiative, MEXT, Japan.  SH also gratefully acknowledges
support by a Grant-in-Aid for Scientific Research (22740153) from the
Japan Society for Promotion of Science (JSPS).

\FloatBarrier

\appendix

\section{The fluxtrap deformation of a set of NS--fivebranes}
\label{sec:ns5-fluxtrap}

\FloatBarrier

\begin{table}
  \centering
  \begin{tabular}{lcccccccccc}
    \toprule
    direction & 0 & 1 & 2 & 3 & 4 & 5 & 6 & 7 & 8 & 9 \\  \midrule 
    NS5 & $\times $ & $\times $ & & & & & $\times $ & $\times $ & $\times $ & $\times$ \\
    fluxtrap & $\times $ & $\times $ & $\times $ & $\times $ & & & & & & $\times$ \\
    D2 & $\times $ & $\times $ & $\times $ \\ 
    D4 & $\times $ & $\times $ & & $\times $ & $\times $ & $\times $ \\ \bottomrule
  \end{tabular}
  \caption{Embedding of the D2--brane with respect to the
    NS5 fluxtrap.}
  \label{tab:NS5-embedding}
\end{table}

\subsection{Bulk fields}

In order precisely to
specify the brane configuration described in outline in section \ref{sec:gaugebethe} of this paper,
we need to consider D2--branes stretched
between parallel NS5--branes.  Consider a stack of parallel NS5--branes
in flat space, extended in the directions $x^1, x^6, x^7, x^8, x^9$
(see Table~\ref{tab:NS5-embedding}). Since the configuration preserves
rotations in the $45$ and $67$ planes it is possible to repeat the
same fluxbrane construction as in Section~\ref{sec:fluxtrap}. The
fields in the bulk in the non--trivial directions read:
\begin{align}
    &\wt {\di s}^2 = U \left[ \di x_2^2 + \di x_3^2 + \di \rho_1^2 + \rho_1^2 \left( \di
      \phi_1 + m \wt R \di \wt u \right)^2
    \right] + \di \rho_2^2 
    + \rho_2^2 \left( \di \phi_2 - m \wt R \di \wt u \right)^2 +
    \wt R^2 \di \wt u^2 \, , \\
  & B = b_i \di x^i \wedge
  \left( \di \phi_1 + m \wt R \di \wt u \right) \, , \\
  & \Phi =\log(\wt R\,g_3^2)+ \frac{1}{2}\log U \, ,
\end{align}
where
\begin{align}
  U &= 1 + \frac{N_5 \alpha'}{x_2^2 + x_3^2 + \rho_1^2} \, , & b_i \di
  x^i &=  \frac{\di U}{\di x_3} \left( - \left( x_3^2 + \rho_1^2 \right)
    \di x_2 + x_2 x_3 \di x_3 + x_2 \rho_1 \di \rho_1 \right) \, ,
\end{align}
so that
\begin{equation}
  \di (b_i \di x^i \wedge \di \theta_1 ) = * \di U \, , 
\end{equation}
where the Hodge star is understood in the four-dimensional space
$(x_2, x_3, \rho_1, \theta_1 )$.

In rectilinear coordinates ($x_4 + \imath x_5 = \rho_1 \eu^{\imath
  \phi_1}, x_6 + \imath x_7 = \rho_2 \eu^{\imath \phi_2}$):
\begin{equation}
  \wt {\di s}^2 = U \left[ \di x_2^2 + \di x_3^2 + \sum_{i=4}^5 \left(
    \di x_i + m V^i \di x_8 \right)^2 \right] + \sum_{i=6}^7 \left(
    \di x_i + m V^i \di x_8 \right)^2 + \di x_8^2 \, ,
\end{equation}
where $V^i \del_i$ is the same vector as in Equation~(\ref{eq:KillVec}):
\begin{equation}
  V^i \del_i = - x^5 \del_{x_4} + x^4 \del_{x_5} + x^7 \del_{x_6} -
  x^6 \del_{x_7} = \del_{\phi_1} - \del_{\phi_2} \, .
\end{equation}
This provides the \emph{$\Omega$--deformation of the NS5 background}.

\bigskip

Following the same procedure as in Section~\ref{sec:fluxtrap} we can
T--dualize in the direction $\wt u$ and get the NS5 fluxtrap
background:
\begin{align}
  \begin{split}
    \di s^2 &= \di \vec x^2_{0 \dots 1} + U \, \left[ \di x_2^2 + \di x_3^2 + \di
      \rho_1^2 + \rho_1^2 \di \phi_1^2 \right] + \di \rho_2^2 +
    \rho_2^2 \di \phi_2^2 + \di x_9^2 \\
    & \phantom{{}={}} + \frac{1}{\Delta^2} \left[ \left( m \, b_i \di x^i
        + \di x_8 \right)^2 - m^2 \left( U \rho_1^2 \di \phi_1 -
        \rho_2^2 \di \phi_2 \right)^2 \right]
\end{split} \\
  \begin{split}
    B &= \frac{1}{\Delta^2} \left[ b_i \di x^i \wedge
      \left( \di \phi_1 + m^2 \rho_2^2 \left( \di \phi_1 + \di \phi_2 \right) \right)
%
   + m 
   \left( U \, \rho_1^2 \di \phi_1 - \rho_2^2 \di
     \phi_2 \right)
   \wedge \di x_8 \right]\,,
  \end{split} \\
  \eu^{-\Phi} &= \frac{1}{g_3^2 \sqrt{\alpha'}} \frac{\Delta}{\sqrt{U}} 
\, ,
\end{align}
where
\begin{equation}
  \Delta^2 = 1 + m^2 \left( U  \rho_1^2 + \rho_2^2
  \right) 
  \, .
\end{equation}
This configuration preserves \textbf{8~real supercharges} and the
Killing spinors have the following explicit expression:
\begin{equation}
  \boxed{\begin{cases} 
    \epsilon_L = \eu^{-\Phi/8}
    \left( \Id + \Gamma_{11} \right) \proj{NS5}_-
    \proj{flux}_-     \exp [\tfrac{1}{2} \phi_1 \Gamma_{45} +
      \tfrac{1}{2} \phi_2 \Gamma_{67}] \epsilon_0 \\
    \epsilon_R = \eu^{-\Phi/8} \left(
      \Id - \Gamma_{11} \right) \Gamma_u \proj{NS5}_+
    \proj{flux}_-     \exp[ \tfrac{1}{2} \phi_1 \Gamma_{45} +
      \tfrac{1}{2} \phi_2 \Gamma_{67}]  \epsilon_1  \\
  \end{cases}}
\end{equation}
where $\epsilon_0 $ and $\epsilon_1 $ are constant Majorana spinors,
\begin{equation}
  \proj{NS5}_\pm = \tfrac{1}{2} \left( \Id \pm \Gamma_{2345}
  \right) \, ,
\end{equation}
and
\begin{equation}
  \Gamma_u = \frac{m \rho_1
    \sqrt{U}}{\Delta} \Gamma_5 - \frac{m \rho_2}{\Delta} \Gamma_7 +
  \frac{1}{\Delta} \Gamma_8 \, .
\end{equation}

\subsection{Open strings}

\paragraph{D2--brane ansatz}

The dynamics of a D2--brane extended in the $(x_1,x_2) $ can be
studied following the parallel computation in
Section~\ref{sec:D2-in-fluxtrap} in absence of NS5--branes.

In order to construct \textsc{bps} solutions, we start
with the ansatz
\begin{align}
  F_{\alpha \beta} &=0 \, , & x_0 &=\zeta^0 \, , & x_1 &= \zeta^1 \, , & x_2 &=
  x_2(\zeta^0,\zeta^1,\zeta^2) \, , & \phi_1 &= \dphi \, \zeta^0 \, ,
  & \phi_2 &= -\dphi \, \zeta^0 \, .
\end{align}
All the other coordinates are are independent of $\zeta^a$.  This
ansatz does not completely fix the reparametrization invariance
($\zeta\mapsto \wt\zeta(\zeta)$) of the D2--brane. The \emph{e.o.m.} for
$x_2$,
\begin{equation}
  \frac{\del}{\del\zeta^\alpha} \frac{\delta\mathcal{L}} {\delta(\del_\alpha x_2) } = \frac{\delta\mathcal{L}} {\delta x_2}
\end{equation}
is satisfied for any choice of $x_2$. This means that we can fix the
\emph{Diff} invariance by choosing the static gauge,
\begin{equation}
  x_2 = \zeta^2.
\end{equation}
Note that the consistency of this choice is automatic because of the reparametrization
invariance of the DBI action, but still appears nontrivial due to the fact that $\del_{x_2}$ is not
  a Killing vector.

The pullbacks of metric and $B$ field are
\begin{align}
  g_{\alpha \beta} \di \zeta^\alpha  \di \zeta^\beta &= - \frac{\dphi^2 +
    \Delta^2 \left( m^2 - \dphi^2 \right)}{m^2 \Delta^2} \left(
    \di \zeta^0 \right)^2 + \left( \di \zeta^1 \right)^2 + \left( U +
    \Lambda^2 \right) \left( \di \zeta^2 \right)^2 \,, \\
  B_{\alpha \beta} \di \zeta^\alpha  \wedge \di \zeta^\beta &= \frac{\Lambda}{\Delta}
  \frac{\dphi}{m} \di \zeta^0 \wedge \di \zeta^2 \, ,
\end{align}
where
\begin{equation}
  \Lambda = \frac{m}{\Delta}  \left( x_3^2 + \rho_1^2 \right) \frac{\di U}{\di x_3} \, .  
\end{equation}
The bosonic part of the DBI action reads
\begin{equation}
  S = -\mu_2 \int \di^3 \zeta \, \sqrt{ 1 - \frac{1 - \Delta^2 \left( 1 + \Lambda^2 / U  \right)}{m^2} \left( m^2 - \dphi^2 \right)} \, .
\end{equation}
Introducing
%
\begin{equation}
  \Xi^2 = U(\zeta_2, x_3, \rho_1) \rho_1^2 + \rho_2^2 + \left( x_3^2 +
    \rho_1^2 \right)^2 \frac{U_{,3}(\zeta_2, x_3, \rho_1)^2}{U(\zeta_2, x_3, \rho_1)} = \frac{\Delta^2 \left( U + \Lambda^2 \right)}{m^2 U} - \frac{1}{m^2}
\end{equation}
the equations of motion reduce to:
\begin{equation}
  \frac{\delta \mathcal{L}}{\delta X^\sigma} = \frac{1}{4 \pi^2 g_3^2
    (\alpha')^2} \frac{\Xi \left( m^2 - \dphi^2 \right)
    \del_\sigma \Xi}{\sqrt{ 1 + \Xi^2 \left( m^2 - \dphi^2 \right)}}
  = 0 \, , \quad \text{where $X^\sigma = \set{
      x_3, \rho_1, \rho_2, x_8, x_9}$.}
\end{equation}
There are two possibilities to satisfy these equations:
\begin{enumerate}
\item If we require
  \begin{equation}
    \Xi = 0 \, ,
  \end{equation}
  this is equivalent to
  \begin{equation}
    \Delta^2 + \frac{\Delta^2 \Lambda^2}{U} = 1 \, .    
  \end{equation}
  Since $\Delta \ge 1$ and $U$ is non-negative, the condition can
  only be satisfied if
  \begin{gather}
    \Delta^2 = 1 \Rightarrow \rho_1 = \rho_2 = 0\,, \\
    \Lambda = 0 \, .
  \end{gather}
  We will refer to this solution where the D2--brane is localized at
  $\rho_1 = \rho_2 = x_3 = 0 $ as the \emph{static embedding}.
\item If 
  \begin{equation}
    \dphi = \pm m \, .
  \end{equation}
  These are the two branches of the \emph{rotating D2--brane embedding}.  Note
  that just like it was in the absence of NS5--branes, even if we are
  not in a linear approximation the frequency is constant and no
  conditions are imposed on the position of the D2--brane in the other
  transverse directions.
\end{enumerate}

\paragraph{Hamiltonian formalism.}

Let us now verify that the rotating solution satisfies has exactly the
\textsc{bps} energy $\mathcal{H} - \mathcal{H}^{\text{static}} = \abs{m \mathcal{J}}$. The angular momentum density associated to the rotation in the
direction of the Killing vector $V^i \del_i$ is:
\begin{equation}
  \mathcal{J} = V^\rho P_\rho = \frac{\delta \mathcal{L}}{\delta \dphi} =
  - \frac{1}{4 \pi^2 g_3^2
    (\alpha')^2} \frac{\Xi^2 \dphi}{\sqrt{1+ \Xi^2
      \left( m^2 - \dphi^2 \right)}} \, ,
\end{equation}
and the Hamiltonian density reads:
\begin{equation}
  \mathcal{H} = P_\rho \dot X^\rho - \mathcal{L} = -
  \frac{1}{4 \pi^2 g_3^2
    (\alpha')^2} \frac{1+ \Xi^2 m^2 }{\sqrt{1 + \Xi^2
      \left( m^2 - \dphi^2 \right)}} \, .
\end{equation}
It follows that on-shell the relation
\begin{equation}
  \left.  \frac{\mathcal{H} -
      \mathcal{H}^{\text{static}}}{\mathcal{J}} \right|_{\dphi =
    \pm m} = \pm m 
\end{equation}
is satisfied without any extra consistency conditions.

\paragraph{Supersymmetry.}

We can now turn to the construction of the Killing spinors preserved
by the D2--brane embeddings we have found above.

The gamma matrices pulled back to the D2--brane are
\begin{align}
  \hat\Gamma_0 &=\Gamma_0-\Gamma_8+\frac{1}{\Delta}\Gamma_u \, ,\\
  \hat\Gamma_1 &=\Gamma_1 \, ,\\
  \hat\Gamma_2 &=\sqrt U \Gamma_2+\Lambda\Gamma_u \, .
\end{align}
The gamma matrix appearing in the kappa symmetry transformation is
modified by the presence of the $B$ field:
\begin{equation}
  \Gamma_{D2} = \frac{1}{\sqrt{-\det(g+B)}} \left( \Id +
    B^{\alpha \beta} \Gamma_{\alpha \beta} \Gamma_{11} \right) \hat \Gamma_{012}
 = \frac{\Delta}{\sqrt U} \left( -\hat\Gamma_{02} +
    \frac{\Lambda}{\Delta} \Gamma_{11} \right) \hat \Gamma_1 \, .
\end{equation}
The condition for preserving supersymmetry is again $\epsilon_L =
\Gamma_{D2}\epsilon_R$, explicitly:
\begin{equation}
  \left( \Id + \Gamma_{11}  \right) \proj{NS5}_- \proj{flux}_- \epsilon_0 = \Gamma_{D2} \left( \Id - \Gamma_{11}  \right) \Gamma_u \proj{NS5}_+ \proj{flux}_- \epsilon_1 \, .
\end{equation}
Plugging in the explicit expression for $\Gamma_u$, and using the fact that:
\begin{align}
  \Gamma_i \proj{flux}_\pm = \begin{cases}
    \proj{flux}_\mp \Gamma_i & \text{if $i = 4,5,6,7$} \\
    \proj{flux}_\pm \Gamma_i & \text{otherwise}
  \end{cases} &&
   \Gamma_i \proj{NS5}_\pm = \begin{cases}
    \proj{NS5}_\mp \Gamma_i & \text{if $i = 2,3,4,5$} \\
    \proj{NS5}_\pm \Gamma_i & \text{otherwise}
  \end{cases}
\end{align}
we find
that the conditions for the preservation of supersymmetry become:
\begin{align}
  \epsilon_0 - \Gamma_{0128} \, \epsilon_1 &= 0\,, \\
  m \rho_1 U \, \Gamma_{25} \left( \Gamma_0 \mp \Gamma_{8} \right)
  \epsilon_1 &= 0\,,  \\
  m \rho_2 \sqrt U  \, \Gamma_{27} \left( \Gamma_0 \mp \Gamma_{8} \right)
  \epsilon_1 &= 0\,, \\
  \Lambda \, \Gamma_1 \left( \Gamma_0 \mp \Gamma_{8} \right) \epsilon_1 &= 0\,.
\end{align}
Again we have two possibilities:
\begin{enumerate}
\item In the static embedding case we have $\rho_1 =
  \rho_2 = \Lambda = 0 $, so we only need to impose the condition
  \begin{equation}
    \epsilon_0 = \Gamma_{0128} \, \epsilon_1 \, .
  \end{equation}
  We find that the \textsc{bps} static brane embedding preserves
  \textbf{4~real supercharges}:
  \begin{equation}
    \boxed{\begin{cases}
      \epsilon_L = \eu^{-\Phi/8} \left( \Id + \Gamma_{11} \right)
      \proj{NS5}_- \proj{flux}_-  \Gamma_{1208} \exp [ \tfrac{1}{2} \left( \phi_1 +
        \phi_2 \right) \Gamma_{67} ] \epsilon_2 \, , \\
      \epsilon_R = \eu^{-\Phi/8} \left( \Id - \Gamma_{11} \right) \Gamma_u
      \proj{NS5}_+ \proj{flux}_-  \exp [ \tfrac{1}{2} \left( \phi_1 +
        \phi_2 \right) \Gamma_{67} ] \epsilon_2 \, .
    \end{cases}}    
  \end{equation}
\item For the rotating embedding $\dphi = \pm m$, together with
  $\epsilon_0 = \Gamma_{1208} \, \epsilon_1 $ we need to impose the extra
  condition
  \begin{equation}
    \epsilon_1 =  \left( \Id \mp \Gamma_{08} \right) \epsilon_2 \, ,
  \end{equation}
  where $\epsilon_2$ is a constant Majorana spinor.  This breaks
  another half of the supersymmetries so that the \textsc{bps}
  rotating brane embedding preserves a total of \textbf{2~real
    supercharges}:
  \begin{equation}
    \boxed{\begin{cases}
      \epsilon_L = \eu^{-\Phi/8} \left( \Id + \Gamma_{11} \right)
      \proj{NS5}_- \proj{flux}_-  \Gamma_{1208} \exp [ \tfrac{1}{2} \left( \phi_1 +
        \phi_2 \right) \Gamma_{67} ] \left( \Id \mp \Gamma_{08} \right) \epsilon_2 \, , \\
      \epsilon_R = \eu^{-\Phi/8} \left( \Id - \Gamma_{11} \right) \Gamma_u 
      \proj{NS5}_+ \proj{flux}_-  \exp [ \tfrac{1}{2} \left( \phi_1 +
        \phi_2 \right) \Gamma_{67} ] \left( \Id \mp \Gamma_{08} \right) \epsilon_2 \, .
    \end{cases}}    
  \end{equation}
\end{enumerate}

\section{Conventions}
\label{sec:conventions}

\FloatBarrier

In this appendix we collect the conventions used in the paper.

\begin{table}
  \centering
  \begin{tabular}{lcc}
    \toprule
    description & index & range \\ \midrule
    spacetime flat & $m, n, \dots $ & $0, \dots, 9 $ \\ 
    spacetime curved & $\mu, \nu, \dots $ & $0, \dots, 9 $ \\
    D2--brane worldvolume & $\alpha, \beta, \dots $ & $0, 1, 2$ \\
    transverse to the D2--brane & $\rho, \sigma, \dots $ & $3, \dots, 9$ \\
    spinor components & $ a,b, \dots $ & $1, \dots, 16 $ \\
    D2--brane-preserved Killing spinors & $ A, B, \dots $ & $1, \dots, 8 $ \\
    \bottomrule
  \end{tabular}
  \caption{Conventions for the choice of indices.}
  \label{tab:indices}
\end{table}

The indices are used according to Table~\ref{tab:indices}. The
signature of the metric is $(-, +, \dots , +)$. Hence the flat Gamma
matrices $\Gamma_m$ satisfy the Clifford algebra:
\begin{equation}
  \{ \Gamma_m , \Gamma_n \} = 2 \, \eta_{mn} = 2 \, \mathrm{diag} \{ -1, 1, \dots,
    1 \} \, .  
\end{equation}
The chirality matrix $\Gamma_{11} $ is given by
\begin{equation}
  \Gamma_{11} = \Gamma_0 \Gamma_1 \cdots \Gamma_9 \, .
\end{equation}
The antisymmetric product of $N$ gamma matrices is normalized as
follows:
\begin{equation}
  \Gamma_{m_1 \dots m_{N}} = \frac{1}{N!}
  \left( \Gamma_{m_1} \dots \Gamma_{m_N}  \pm \text{permutations} \right) \, .
\end{equation}
The gamma matrices in the bulk are
\begin{equation}
  \Gamma_\mu = e\ud{m}{\mu} \Gamma_m \, ,\hspace{2em} \{ \Gamma_\mu,
  \Gamma_\nu \} = g_{\mu \nu} \, ,
\end{equation}
and their pullbacks on the D--brane are given by
\begin{equation}
  \Gamma_\alpha = \frac{\del X^\mu}{\del \zeta^\alpha } e\ud{m}{\mu}
  \Gamma_m \, ,\hspace{2em} \{ \Gamma_\alpha, \Gamma_\beta \} = g_{\alpha
    \beta} \, . 
\end{equation}
In order to avoid confusion, the pullback of the gamma matrices in the
$\zeta^0$ direction is denoted by $\hat \Gamma_0$:
\begin{equation}
  \hat \Gamma_0 = \left. \Gamma_\alpha \right|_{\alpha = 0} = \frac{\del X^\mu
  }{\del \zeta^0} e\ud{m}{\mu} \Gamma_m \, . 
\end{equation}

\bigskip

In type IIA the spinors are Majorana and are decomposed into the
sum of two chiral components:
\begin{equation}
  \epsilon = \epsilon_L + \epsilon_R \, , \hspace{2em} \Gamma_{11}
  \epsilon_{L} =  \epsilon_{L} \, , \hspace{2em} \Gamma_{11}
  \epsilon_{R} = - \epsilon_{R} \, .  
\end{equation}
The conjugate is defined as:
\begin{equation}
  \bar \epsilon = \imath \epsilon^T \Gamma^0 = - \imath \epsilon^T
  \Gamma_0 \, . 
\end{equation}

\bigskip
\noindent
The supersymmetry transformations of the dilatino $\lambda$ and the
gravitino $\Psi_m$ are given \emph{e.g.} in~\cite{Schwarz:1983qr}:
\begin{align}
\label{eq:delta-dilatino}
  \delta_\epsilon \lambda &= \frac{\eu^{\Phi/4}}{\sqrt{2}} \left[-\frac{1}{2}
    \del_m \, \Phi \, \Gamma^m \Gamma_{11} + \frac{1}{24}\, H_{mnp}\,
    \Gamma^{mnp} \right]\epsilon \,, \\
\label{eq:delta-gravitino}
  \delta_\epsilon \Psi_m &= \eu^{ \Phi / 4} \left[ \nabla_m + \frac{1}{8}\,
    \del_n \Phi \,\Gamma\ud{n}{m} + \frac{1}{96}\, H_{npq} \, \left(\,
      \Gamma\du{m}{npq}- 9\, \delta\du{m}{n} \,\Gamma^{pq} \, \right)\,
    \Gamma_{11} \right] \epsilon  \,,
\end{align}
where the action of the covariant derivative on a spinor is given by
\begin{equation}
  \nabla_m\, \epsilon = \del_m\, \epsilon + \frac{1}{4}
  \omega\du{m}{np} \Gamma_{np}\, \epsilon\,,
\end{equation}
and $\omega $ is the spin connection. %
A spinor $\epsilon$ is a Killing spinor if
\begin{align}
  \delta_\epsilon \lambda &= 0 & \text{and} && \delta_\epsilon \Psi_m &= 0 \, .
\end{align}

\newpage

\bibliography{OmegaString}

\providecommand{\href}[2]{#2}\begingroup\raggedright\begin{thebibliography}{10}

\bibitem{Melvin:1963qx}
M.~A. Melvin, {\it {Pure magnetic and electric geons}},  {\em Phys. Lett.} {\bf
  8} (1964) 65--70.

\bibitem{Tseytlin:1994ei}
A.~A. Tseytlin, {\it {Melvin solution in string theory}},  {\em Phys.Lett.}
  {\bf B346} (1995) 55--62, [\href{http://xxx.lanl.gov/abs/hep-th/9411198}{{\tt
  hep-th/9411198}}].

\bibitem{Tseytlin:1995zv}
A.~A. Tseytlin, {\it {Closed superstrings in magnetic field: Instabilities and
  supersymmetry breaking}},  {\em Nucl.Phys.Proc.Suppl.} {\bf 49} (1996)
  338--349, [\href{http://xxx.lanl.gov/abs/hep-th/9510041}{{\tt
  hep-th/9510041}}].

\bibitem{Russo:2001na}
J.~Russo and A.~A. Tseytlin, {\it {Supersymmetric fluxbrane intersections and
  closed string tachyons}},  {\em JHEP} {\bf 0111} (2001) 065,
  [\href{http://xxx.lanl.gov/abs/hep-th/0110107}{{\tt hep-th/0110107}}].

\bibitem{Gutperle:2001mb}
M.~Gutperle and A.~Strominger, {\it {Fluxbranes in string theory}},  {\em JHEP}
  {\bf 0106} (2001) 035, [\href{http://xxx.lanl.gov/abs/hep-th/0104136}{{\tt
  hep-th/0104136}}].

\bibitem{Takayanagi:2001jj}
T.~Takayanagi and T.~Uesugi, {\it {Orbifolds as Melvin geometry}},  {\em JHEP}
  {\bf 0112} (2001) 004, [\href{http://xxx.lanl.gov/abs/hep-th/0110099}{{\tt
  hep-th/0110099}}].

\bibitem{Hellerman:2006tx}
S.~Hellerman and J.~Walcher, {\it {Worldsheet CFTs for Flat Monodrofolds}},
  \href{http://xxx.lanl.gov/abs/hep-th/0604191}{{\tt hep-th/0604191}}.

\bibitem{Moore:1997dj}
G.~W. Moore, N.~Nekrasov, and S.~Shatashvili, {\it {Integrating over Higgs
  branches}},  {\em Commun.Math.Phys.} {\bf 209} (2000) 97--121,
  [\href{http://xxx.lanl.gov/abs/hep-th/9712241}{{\tt hep-th/9712241}}].

\bibitem{Lossev:1997bz}
A.~Lossev, N.~Nekrasov, and S.~L. Shatashvili, {\it {Testing Seiberg-Witten
  solution}},  \href{http://xxx.lanl.gov/abs/hep-th/9801061}{{\tt
  hep-th/9801061}}.

\bibitem{Nekrasov:2002qd}
N.~A. Nekrasov, {\it {Seiberg-Witten prepotential from instanton counting}},
  {\em Adv.Theor.Math.Phys.} {\bf 7} (2004) 831--864,
  [\href{http://xxx.lanl.gov/abs/hep-th/0206161}{{\tt hep-th/0206161}}]. To
  Arkady Vainshtein on his 60th anniversary.

\bibitem{Nekrasov:2003rj}
N.~Nekrasov and A.~Okounkov, {\it {Seiberg-Witten theory and random
  partitions}},  \href{http://xxx.lanl.gov/abs/hep-th/0306238}{{\tt
  hep-th/0306238}}.

\bibitem{Billo:2006jm}
M.~Billo, M.~Frau, F.~Fucito, and A.~Lerda, {\it {Instanton calculus in R-R
  background and the topological string}},  {\em JHEP} {\bf 0611} (2006) 012,
  [\href{http://xxx.lanl.gov/abs/hep-th/0606013}{{\tt hep-th/0606013}}].

\bibitem{Nekrasov:2010ka}
N.~Nekrasov and E.~Witten, {\it {The Omega Deformation, Branes, Integrability,
  and {L}iouville Theory}},  \href{http://xxx.lanl.gov/abs/1002.0888}{{\tt
  arXiv:1002.0888}}.

\bibitem{Nekrasov:2009uh}
N.~A. Nekrasov and S.~L. Shatashvili, {\it {Supersymmetric vacua and {B}ethe
  ansatz}},  {\em Nucl. Phys. Proc. Suppl.} {\bf 192-193} (2009) 91--112,
  [\href{http://xxx.lanl.gov/abs/0901.4744}{{\tt arXiv:0901.4744}}].

\bibitem{Nekrasov:2009ui}
N.~A. Nekrasov and S.~L. Shatashvili, {\it {Quantum integrability and
  supersymmetric vacua}},  {\em Prog. Theor. Phys. Suppl.} {\bf 177} (2009)
  105--119, [\href{http://xxx.lanl.gov/abs/0901.4748}{{\tt arXiv:0901.4748}}].

\bibitem{Orlando:2010uu}
D.~Orlando and S.~Reffert, {\it {Relating Gauge Theories via Gauge/Bethe
  Correspondence}},  {\em JHEP} {\bf 1010} (2010) 071,
  [\href{http://xxx.lanl.gov/abs/1005.4445}{{\tt arXiv:1005.4445}}].

\bibitem{Hanany:1997vm}
A.~Hanany and K.~Hori, {\it {Branes and N = 2 theories in two dimensions}},
  {\em Nucl. Phys.} {\bf B513} (1998) 119--174,
  [\href{http://xxx.lanl.gov/abs/hep-th/9707192}{{\tt hep-th/9707192}}].

\bibitem{deBoer:1997ka}
J.~de~Boer, K.~Hori, Y.~Oz, and Z.~Yin, {\it {Branes and mirror symmetry in N=2
  supersymmetric gauge theories in three-dimensions}},  {\em Nucl.Phys.} {\bf
  B502} (1997) 107--124, [\href{http://xxx.lanl.gov/abs/hep-th/9702154}{{\tt
  hep-th/9702154}}].

\bibitem{deBoer:1997kr}
J.~de~Boer, K.~Hori, and Y.~Oz, {\it {Dynamics of N=2 supersymmetric gauge
  theories in three-dimensions}},  {\em Nucl.Phys.} {\bf B500} (1997) 163--191,
  [\href{http://xxx.lanl.gov/abs/hep-th/9703100}{{\tt hep-th/9703100}}].

\bibitem{Aharony:1997bx}
O.~Aharony, A.~Hanany, K.~A. Intriligator, N.~Seiberg, and M.~Strassler, {\it
  {Aspects of N=2 supersymmetric gauge theories in three-dimensions}},  {\em
  Nucl.Phys.} {\bf B499} (1997) 67--99,
  [\href{http://xxx.lanl.gov/abs/hep-th/9703110}{{\tt hep-th/9703110}}].

\bibitem{Polchinski:1995mt}
J.~Polchinski, {\it {Dirichlet Branes and Ramond-Ramond charges}},  {\em
  Phys.Rev.Lett.} {\bf 75} (1995) 4724--4727,
  [\href{http://xxx.lanl.gov/abs/hep-th/9510017}{{\tt hep-th/9510017}}].

\bibitem{Aharony:2008ug}
O.~Aharony, O.~Bergman, D.~L. Jafferis, and J.~Maldacena, {\it {N=6
  superconformal Chern-Simons-matter theories, M2-branes and their gravity
  duals}},  {\em JHEP} {\bf 0810} (2008) 091,
  [\href{http://xxx.lanl.gov/abs/0806.1218}{{\tt arXiv:0806.1218}}].

\bibitem{Buscher:1987sk}
T.~Buscher, {\it {A Symmetry of the String Background Field Equations}},  {\em
  Phys.Lett.} {\bf B194} (1987) 59.

\bibitem{Bakas:1995hc}
I.~Bakas and K.~Sfetsos, {\it {T duality and world sheet supersymmetry}},  {\em
  Phys.Lett.} {\bf B349} (1995) 448--457,
  [\href{http://xxx.lanl.gov/abs/hep-th/9502065}{{\tt hep-th/9502065}}].

\bibitem{Bergshoeff:1994cb}
E.~Bergshoeff, R.~Kallosh, and T.~Ortin, {\it {Duality versus supersymmetry and
  compactification}},  {\em Phys.Rev.} {\bf D51} (1995) 3009--3016,
  [\href{http://xxx.lanl.gov/abs/hep-th/9410230}{{\tt hep-th/9410230}}].

\bibitem{Martucci:2005rb}
L.~Martucci, J.~Rosseel, D.~Van~den Bleeken, and A.~Van~Proeyen, {\it {Dirac
  actions for D-branes on backgrounds with fluxes}},  {\em Class.Quant.Grav.}
  {\bf 22} (2005) 2745--2764,
  [\href{http://xxx.lanl.gov/abs/hep-th/0504041}{{\tt hep-th/0504041}}].

\bibitem{Bergshoeff:1996tu}
E.~Bergshoeff and P.~Townsend, {\it {Super D-branes}},  {\em Nucl.Phys.} {\bf
  B490} (1997) 145--162, [\href{http://xxx.lanl.gov/abs/hep-th/9611173}{{\tt
  hep-th/9611173}}].

\bibitem{Brodie:1997wn}
J.~H. Brodie, {\it {Two dimensional mirror symmetry from M-theory}},  {\em
  Nucl. Phys.} {\bf B517} (1998) 36--52,
  [\href{http://xxx.lanl.gov/abs/hep-th/9709228}{{\tt hep-th/9709228}}].

\bibitem{Witten:1997ep}
E.~Witten, {\it {Branes and the dynamics of QCD}},  {\em Nucl.Phys.} {\bf B507}
  (1997) 658--690, [\href{http://xxx.lanl.gov/abs/hep-th/9706109}{{\tt
  hep-th/9706109}}].

\bibitem{Hanany:1996ie}
A.~Hanany and E.~Witten, {\it {Type IIB superstrings, BPS monopoles, and three-
  dimensional gauge dynamics}},  {\em Nucl. Phys.} {\bf B492} (1997) 152--190,
  [\href{http://xxx.lanl.gov/abs/hep-th/9611230}{{\tt hep-th/9611230}}].

\bibitem{Orlando:2010aj}
D.~Orlando and S.~Reffert, {\it {The Gauge-Bethe Correspondence and Geometric
  Representation Theory}},  \href{http://xxx.lanl.gov/abs/1011.6120}{{\tt
  arXiv:1011.6120}}.

\bibitem{Seiberg:1994rs}
N.~Seiberg and E.~Witten, {\it {Electric - magnetic duality, monopole
  condensation, and confinement in N=2 supersymmetric Yang-Mills theory}},
  {\em Nucl.Phys.} {\bf B426} (1994) 19--52,
  [\href{http://xxx.lanl.gov/abs/hep-th/9407087}{{\tt hep-th/9407087}}].

\bibitem{DomenicoTalk}
D.~Orlando, {\it The gauge-{B}ethe correspondence: Dualities and branes --
  {T}alk given at the {B}ranes and {B}ethe {A}nsatz in supersymmetric gauge
  theories {W}orkshop, {S}imons {C}enter for {G}eometry and {P}hysics, march
  2011}, .

\bibitem{Schwarz:1983qr}
J.~H. Schwarz, {\it {Covariant Field Equations of Chiral N=2 D=10
  Supergravity}},  {\em Nucl.Phys.} {\bf B226} (1983) 269.

\end{thebibliography}\endgroup

\end{document}